\documentclass[aps,prd,a4paper,showpacs,12pt]{article}

\usepackage{multicol}
\usepackage{subfigure}
\usepackage {graphicx}
\usepackage{color}
\usepackage{xcolor}
\usepackage{threeparttable}
\usepackage{amsfonts}
\usepackage{times}
\usepackage{setspace}
\usepackage{balance}
\usepackage{lastpage}
\usepackage{amsthm}
\usepackage[tbtags]{amsmath}
\usepackage{slashed}
\usepackage{nomencl}
\usepackage{pstricks}
\usepackage{pstricks-add}
\usepackage{pst-plot,pstricks-add}

\usepackage{amsmath,amsfonts,amssymb,simplewick}
\usepackage{float}
\usepackage{amsmath,amssymb,amsfonts,epsfig,graphicx,euscript}%
\usepackage{hyperref}
\hypersetup{bookmarks=true, 
bookmarksnumbered=true,
linktoc=page, 
pdfstartview={FitH}, 
colorlinks=true, 
citecolor=red, 
filecolor=blue, 
linkcolor=blue, 
urlcolor=blue}
\usepackage{amsmath}
\usepackage{amsfonts}
\usepackage{graphicx}
\usepackage{caption}
\usepackage{float}
\usepackage[all]{hypcap}
\numberwithin{equation}{section}
\usepackage{cite}
\usepackage{calc}
\usepackage{mathtools}
\newlength{\spacer}
\newsavebox{\mybox}

\DeclareMathOperator{\Tr}{Tr}
\DeclareMathOperator{\Det}{Det}
\DeclareMathOperator{\Ln}{Ln}

\DeclareMathOperator{\csch}{csch}

\DeclareMathAlphabet{\mathpzc}{OT1}{pzc}{m}{it}
\usepackage{ragged2e}
\renewcommand{\thefootnote}{\fnsymbol{footnote}}

\begin{document}
	\begin{center}
		{\large{\textbf{Casimir free energy for massive fermions: a comparative study of various approaches}}}
		\vspace*{1.5cm}
		\begin{center}
			{\bf M. Sasanpour\footnote{m\_sasanpour@sbu.ac.ir}, C. Ajilyan\footnote{chn.ajilian@gmail.com} and S. S. Gousheh\footnote{ss-gousheh@sbu.ac.ir}}\\
		\vspace*{0.5cm}
		{\it{Department of Physics, Shahid Beheshti University, Tehran, Iran}}\\
		\vspace*{1cm}
	\end{center}
\end{center}
	\begin{center}
		\today
	\end{center}

\renewcommand*{\thefootnote}{\arabic{footnote}}
\setcounter{footnote}{0}
\date{\today}
\textbf{Abstract:}
We compute the Casimir thermodynamic quantities for a massive fermion field between two parallel plates with the MIT boundary conditions, using three different general approaches and present explicit solutions for each. The Casimir thermodynamic quantities include the Casimir Helmholtz free energy, pressure, energy and entropy. The three general approaches that we use are based on the fundamental definition of Casimir thermodynamic quantities, the analytic continuation method represented by the zeta function method, and the zero temperature subtraction method. We include the renormalized versions of the latter two approaches as well, whereas the first approach does not require one. Within each general approach, we obtain the same results in a few different ways to ascertain the selected cancellations of infinities have been done correctly. We then do a comparative study of the three different general approaches and their results, and show that they are in principle not equivalent to each other and they yield, in general, different results. In particular, we show that the Casimir thermodynamic quantities calculated only by the first approach have all three properties of going to zero as the temperature, the mass of the field, or the distance between the plates increases.

\vspace*{25pt}


%
%
%

\section {Introduction}
\indent

In 1948, Casimir~\cite{r1Cas.} predicted an attractive force between neutral parallel conducting plates due to the zero-point energy of the quantized electromagnetic field. The Casimir effects are among the most interesting manifestations of the vacuum in quantum field theory, and have been studied extensively for more than 70 years. These effects arise when a system is subject to nonperturbative conditions or constraints, such as boundary conditions, background fields such as solitons, or nontrivial space-time backgrounds. These nonperturbative constraints are part of the definition of the system, including its vacuum. The Casimir effects have many applications in various branches of physics such as particle physics~\cite{r32Particle., r322Particle., r323Particle., r324Particle.}, condensed matter and laser physics~\cite{r33Cond., r332Cond., r333Cond., r334Cond.}, nanotechnology~\cite{r33Nano., r332Nano., r333Nano., r334Nano.}, string theory~\cite{r34String., r342String., r343String.}, and cosmology~\cite{r35Cosmo., r352Cosmo., r353Cosmo., r354Cosmo.}. In connection with the experimental aspect, Sparnaay was the first to investigate the Casimir force~\cite{r3Spar.}, but Lamoreaux et al.~\cite{r4Lamo.} were the first to measure it with acceptable precision. For a comprehensive review, see for example~\cite{r32Particle., r30Kimb., r302Kimb., r31Bord2.}. 

In this paper, we explore the differences between three of the commonly used general approaches for calculating the finite temperature Casimir effects. Our reference general approach is based on the fundamental definition of the Casimir thermodynamic quantities which for the Casimir Helmholtz free energy, for example, is the difference between the infinite vacuum Helmholtz free energies of systems subject to the constraints and the corresponding ones that are free from them, both being at the same temperature. We shall henceforth refer to this as the fundamental approach. The other two approaches are based on the zeta function method and the zero temperature subtraction method, respectively. We also include the renormalized versions of the latter two approaches, and shall refer to them collectively as the zeta function approach (ZFA), and the zero temperature subtraction approach (ZTSA), respectively. A more appropriate name for ZFA is the analytic continuation approach, in which the zeta function is the most practical utility. As is well known, the result of analytic continuations is unique, and one of the questions that we want to address here is whether this unique result is the physically acceptable one we seek.

In order to be concrete, we find it necessary to concentrate on an illustrative example which we choose to be a massive fermion between two parallel plates with the MIT boundary condition. In this paper, we present calculations for Casimir thermodynamic quantities within each of the general approaches mentioned above, and present their results in explicit forms which, as far as we know, have not been presented heretofore. Moreover, to ascertain the validity of our results, we present a few different ways of obtaining the same results within each general approach. We then do a comparative study of the three general approaches and their results. Before we start with the computations, we briefly review the historical development of the finite temperature Casimir effects and the use of various approaches.

The fundamental definition of the zero temperature Casimir energy, as stated by Casimir in 1948, is the difference between the zero point energies of the system with and without the constraints. Finite temperature Casimir effect was first introduced by Lifshitz~\cite{r2Lif.} in 1956, who calculated the attractive force between two parallel dielectric plates, at finite temperature by introducing fluctuating electromagnetic field. At high temperatures, the Casimir pressure was found to be proportional to the temperature. This term was later denoted as the classical term. Later on, Mehra~\cite{r5Mehr.} in 1967, used the Helmholtz free energy, which we shall henceforth refer to simply as the free energy, to calculate the thermal correction to the zero temperature Casimir pressure for a conducting cubic cavity. In that paper, the Casimir pressure was calculated as the difference between the pressure inside and outside of the cube, both being at the same temperature. His results also included the classical term at high temperatures.

The next major work on thermal corrections is due to Brown and Maclay~\cite{r6Brown.} in 1969, who calculated the electromagnetic stress-energy tensor between two conducting parallel plates. Using the image-source construction, they obtained the components of the tensor as thermodynamic variables, without any divergent terms. However, for the first time, the final results for both the Casimir pressure and energy density included terms due to the black-body radiation which are proportional to $T^4$. 

In a series of papers from 1976 to 1980, Dowker et al.~\cite{r8Dowk76., r8Dowk78., r8Dowk80.} calculated the vacuum expectation value of the stress-energy tensor at finite temperature using the Green function formalism for a scalar field in curved space-time. They used three different renormalization schemes to obtain finite results. First, they subtracted the $(0,0)$ temperature-spatial mode. Second, they used a `Casimir renormalization' as the difference between free energies before and after constructing the boundary, both being at the same temperature, to compute the heat kernel coefficients. This is analogous to the fundamental approach. Third, they subtracted the contribution of the free Green function at the zero temperature, which they referred to as `the standard flat space renormalization prescription'. This is equivalent to ZTSA. The high temperature limit of $\langle T_{00}\rangle $ in their first and third work had terms proportional to $T^4$ and $T$, while Casimir free energy in their second work had terms proportional to $T^3$, $T$ and $T \ln T$. 

In 1978, Balian and Duplantier~\cite{r9Balian.} defined and used the fundamental definition of the Casimir free energy for the electromagnetic field in a region bounded by thin perfect conductors with arbitrary smooth shapes. The high temperature limit of their results for parallel plates was proportional to $T$, while for the enclosures included an additional term proportional to $T\ln T$.

In 1983, Ambj$\o$rn and Wolfram~\cite{r10Wolf.} computed the Casimir energy and entropy for scalar and electromagnetic fields in a hypercuboidal region, using the generalized zeta function along with the reflection formula as an analytic continuation technique. They showed that the high temperature limit of the Casimir energy for the scalar field in a rectangular cavity in $3+1$ dimensions includes terms proportional to $T^4$, $T^2$ and $T\ln T$. In 1991, Kirsten~\cite{r12Kris.} computed the heat kernel coefficients for the grand thermodynamic potential for a massive bosonic field in hypercuboids in n-dimensions subject to the Dirichlet boundary condition, using the zeta function, and in four dimensions obtained terms proportional to $T^4$, $T^3$, $T^2$, $T$ and $T \ln T$. 

In 1992, Elizalde and Romeo~\cite{r51Elizald92.} calculated only the high and low temperatures expansions of the free energy for a massive bosonic field in hypercuboids of arbitrary dimensions, using multidimensional Epstein zeta functions. They indicated that, as stated in~\cite{r11Plunien.}, to calculate the Casimir free energy, one has to subtract the free energy of the unconstrained boson field, which would eliminate only the $T^4$ term at high temperatures. In 2008, Geyer et al.~\cite{r17Geyer.} suggested a renormalization procedure to calculate the finite temperature free energy, which would supplement the use of zeta function. They stated that the use of zeta function does not include all necessary subtractions, and the terms proportional to powers of $T$ higher than the classical terms obtained in the high temperature limit from the heat kernel method, have to be subtracted. Subsequently, in 2009, Bordag et al.~\cite{r31Bord2.} presented a general picture of the renormalization for the Casimir free energy within the ZFA. They used the heat kernel coefficients for subtraction of the extra terms at low and high temperature limits.

So far, we have mentioned some of the controversies over the finite temperature Casimir effects for bosons. Here, for illustrative purposes, we focus on the finite temperature Casimir effects for the fermion fields. This subject started with the work of Gundersen and Ravndal in 1988~\cite{r22Rav.}, where they explicitly calculated the Casimir free energy for a massless fermion field at finite temperature between two parallel plates. They defined the Casimir free energy as the difference between the free energy in the presence of the plates at temperature $T$ and that of the free space at $T=0$. We have referred to this approach as the zero temperature subtraction approach (ZTSA). The results that they obtained for the Casimir free energy and pressure included $T^4$ terms, with the force being attractive at low temperatures, and repulsive and increasing without bound at high temperatures.

In 2004, Santana et al.~\cite{r25Santana.} calculated the Casimir pressure and energy by a generalized Bogoliubov transformation for a massless fermion field in the cases of two parallel plates, square wave-guides, and cubic boxes, and they confirmed the results of~\cite{r22Rav.}. They used the zeta function as a supplementary part of their calculation. The high temperature limits of these results contain the $T^4$ terms which are equivalent to the black-body terms. Since then, the ZFA has been employed in some papers for fermions, yielding $T^4$ for the high temperature limit of the Casimir effects~\cite{r28Flach., r282Flach., r283Flach.}.

In 2010, Cheng~\cite{r26Cheng.} calculated the Casimir force for a massless Majorana fermion field between two parallel plates using the piston approach. He used the zeta function analytic continuation for both parts of the piston, and upon subtracting the two forces, he obtained a Casimir force which is always attractive and approaches zero as temperature increases. In 2011, Khoo and Teo~\cite{r27Khoo.} presented a similar analysis for massive fermions with extra compact dimensions, and found that the Casimir force is always attractive at any temperature. Also, they stated that the high temperature limits of their results for the Casimir free energy and force contain a term proportional to $T$, which they called the classical term. In 2018, Mo and Jia~\cite{r29Junji.} considered a massless fermion field confined in a rectangular box and defined a renormalized free energy by subtracting the free black-body term along with possible terms proportional to $T^2$ and $T^3$ so as to eliminate the high temperature divergences, with reference to Geyer's work~\cite{r17Geyer.}. They used the Schl\"{o}milch formula which is based on the zeta function. They showed that, after subtracting these terms, both the Casimir free energy and force for parallel plates go to zero at high temperatures.

As is apparent from the historical outline presented, the zeta function has been used extensively in the calculations of the Casimir effects to evaluate the sums over the regular spatial and Matsubara modes, and often as an analytical continuation technique. In some methods, zeta function is used explicitly to calculate the Casimir thermodynamic quantities, e.g.\ in~\cite{r51Elizald92.}, or implicitly, e.g.\ in the Schl\"{o}milch's formula, or as a supplementary part. Examples of the latter include the heat kernel method and the Bogoliubov transformation where the zeta function is used to evaluate the final summations. As we shall show, the use of the generalized zeta function for the sum over spatial modes of the fermion fields, which is possible only for the massless cases since they are regular, is equivalent to subtracting the case with no boundaries at zero temperature, and this yields the correct results for the zero temperature cases. Furthermore, as we shall show, the use of the zeta function for the sum over Matsubara frequencies, which are always regular, is roughly equivalent to subtracting the zero temperature case\footnote{We shall clarify this statement in Secs.~(\ref{zeta},\ref{Epstein})}. These imply that while the ZFA, as well as the ZTSA, is in principle equivalent to the fundamental approach at zero temperature, these equivalencies deserve further investigations for finite temperatures. A broader question that we want to address is the following: when does the removal of infinities by analytic continuations, whose results are certainly unique, yield the correct finite value as the physical results. The same question can be asked about the subtraction methods, such as ZTSA, whose results, contrary to the analytic continuation, are not unique.


As mentioned before, in this paper, we solve the massive fermionic case between two parallel plates at finite temperatures by three different general approaches, {\it i.e.}, the fundamental approach, the ZFA, and the ZTSA. Within each general approach, we display or outline multiple ways of computing the same physical quantities and use various methods and computational techniques to ascertain the selected and delicate cancellations of divergent sums and integrals have been done correctly, all yielding equivalent results. These methods include the Poison summation formula, the Abel-Plana formula, and the Principle of the Argument theorem. As for the results, there are three limits in which one might naturally expect the Casimir effects to vanish: large plate separation, large mass limit, and the high temperature limit\footnote{These can be considered as normalization conditions within the renormalization program for ZFA~\cite{r31Bord2.}.}.
The latter holds if, contrary to the bosonic case, there are no classical terms proportional to $T$ in Casimir thermodynamic quantities for fermions. As we shall show, this is indeed the case and all of the Casimir thermodynamic quantities obtained using the fundamental approach have all three of the desired properties, {\it i.e.}, the Casimir free energy, pressure, energy, and entropy, go to zero at those limits. Indeed, in this approach, the subtraction of the thermodynamic quantities of the constrained and unconstrained systems at the same temperature yields the correct results, including the mentioned limits, without having to label any terms in the resulting expressions for the Casimir quantities as unphysical, and subsequently removing them by hand. 

On the other hand, as we shall show, the results obtained for the massive case using the ZFA and the ZTSA are not equivalent to those of the fundamental approach, nor are they equivalent to each other. They contain extra nonpolynomial terms in variables $T$ and $m$ which, for example, actually diverge as $T \rightarrow \infty$. For the renormalized versions of the latter two approaches, we obtain the high temperature limits of these extra terms, both directly and by the heat kernel method, as polynomials in $T$, up to and including the black-body term $T^d$ in $d$ space-time dimensions, and subtract them according to the renormalization programs introduced\cite{r17Geyer.,r29Junji.}. However, divergences at high temperatures persist in the form of $\ln (T)$ terms. On the other hand, as we shall show, for the massless case, the extra terms can be removed by the renormalization programs introduced~\cite{r17Geyer.,r29Junji.}, or cancel out in the piston method for the Casimir pressure~\cite{r17Geyer.,r26Cheng.,r27Khoo.}. Hence, for the massless case, the results of the fundamental approach, the ZFA (supplemented with the piston approach or its renormalized version), and the renormalized version of ZTSA are all equivalent. However, as we shall show, the fundamental approach yields the correct results for both the massless and massive cases, without the need for any supplementary renormalization program.

The outline of the paper is as follows. In Sec.\ \ref{Helmholtz free energy}, we present two forms for the free energy for a fermion field at finite temperature, which is subject to the MIT boundary conditions at two plates but is otherwise free, starting with the path integral formalism. In Sec.\ \ref{FCasimir}, we calculate the Casimir free energy for a massless fermion field using the fundamental approach and the Poisson summation formula, and show that the Casimir free energy and pressure go to zero at high temperatures and large plate separations. In Sec.\ \ref{zeta}, we calculate the Casimir free energy and pressure of a massless fermion field using the ZFA and the ZTSA, obtaining identical results which have extra black-body terms as compared to the results based on the fundamental approach. We then show how the renormalization program subtracts these extra black-body terms, yielding the correct results, based on the fundamental approach. In Sec.\ \ref{massive}, we consider a massive fermion field as the simplest nontrivial example, and calculate the Casimir free energy, pressure, energy, and entropy using the fundamental approach. We show that they all go to zero in the high temperature and large mass limits. In Sec.\ \ref{Epstein}, we calculate the Casimir free energy and pressure for the same massive fermion problem as in Sec.\ \ref{massive}, using the other two general approaches, {\it i.e.}, the ZFA and the ZTSA, including their renormalized versions, obtaining four different sets of results none of which is equivalent to that based on the fundamental approach. Moreover, we show that in the high temperature limit, the results of the unrenormalized versions diverge as $T^4$, while those of the renormalized versions diverge as $\ln T$. As a side note, we present the condition under which the piston method would always yield the correct Casimir pressure. Finally, in Sec.\ \ref{summary}, we present our conclusions.

\section {The Helmholtz Free Energy}\label{Helmholtz free energy}
\indent
Historically, the first and most commonly used approach to thermal field theory is the imaginary-time formalism. This approach started with the work of Felix Bloch in 1932, who noticed the analogy between the inverse temperature and imaginary-time~\cite{r36Bloch.}, which led to the so-called temperature Green functions with purely imaginary-time arguments. In 1955, Matsubara presented the first systematic approach to investigate quantum field theory at finite temperature by imaginary-time formalism, using the Wick rotation~\cite{r37Mats.}. The discrete frequencies in this formalism are known as Matsubara frequencies. In 1957, Ezawa et al.\ extended the Matsubara's work to the relativistic quantum field theory~\cite{r38Ezawa.}. They discovered the periodicity (anti-periodicity) conditions for the Green function of boson (fermion) fields, the generalization of which became known as the KMS (Kubo~\cite{r38Kubo.} (1957), Martin and Schwinger~\cite{r38MaSch.} (1959)) condition. In the 1960s, Schwinger~\cite{r38Schwin.}, Keldysh~\cite{r38Keld.}, and others~\cite{r38Mill., r382Mill., r383Mill.} developed the real time formalism for the finite temperature field theory. The latest development of this formalism was presented by Takahashi and Umezawa~\cite{r38Umeza., r382Umeza.}, based on an operator formulation of the field theory at finite temperature, which is called thermofield dynamics (TFD). Since then, many subjects in finite temperature field theory, e.g., thermal Ward-Takahashi relations, KMS relations, renormalization procedure, have been studied and reported in for example~\cite{r40Kapusta., r41Bellac., r41Khanna., r42Lands., r42Lain.}.

In this paper, we use the Matsubara formalism to study the Casimir effect for a free fermion field confined between two parallel plates at finite temperature. In this formalism, a Euclidean field theory is obtained by a Wick rotation on the time coordinate, $t \to  - i \tau$, such that, the Euclidean time $\tau$ is confined to the interval $\tau \in [0 , \beta]$, where $\beta=(kT)^{-1}$ ~\cite{r37Mats., r41Bellac.}. The partition function in the path integral representation becomes:
\begin{equation}\label{s1}
Z = \int\limits_{\begin{array}{l}
\scriptstyle \psi (\beta ,\overrightarrow r ) =  - \psi (0 ,\overrightarrow r)\hfill \\
\scriptstyle \overline \psi  (\beta ,\overrightarrow r ) =  - \overline \psi  (0 ,\overrightarrow r) \hfill 
\end{array}}
 D \overline \psi  D \psi   \exp\left( - \int_{0}^\beta  d\tau  \int {d^3 x} \mathcal{L}_E \right) .
\end{equation}
For a free fermion field, this expression simplifies as follows: 
\begin{eqnarray}\label{s2}
Z &=& \det \left( \gamma _E^{\mu} \partial _{\mu E} + m \right).
\end{eqnarray}
Using the partition function given by Eq.~(\ref{s2}), the free energy is obtained as
\begin{equation}\label{s3}
F =  - \frac{\ln(Z)}{\beta } =  - T \ln \left[ \det \left( \gamma _E^{\mu} \partial _{\mu E} + m \right) \right] = 
- 2T  \Tr \left[ \ln\left( P_E^{2} + m^2 \right) \right] .
\end{equation}
The trace in Eq.~(\ref{s3}) indicates the summation over eigenvalues of Dirac operator in the momentum space representation. Moreover, the modes of zero-component of momentum or the Matsubara frequencies are discrete, due to the KMS anti-periodicity condition on the finite $\tau$ interval: 
\begin{equation}\label{s3a}
 \omega_{n_0} = \frac{n_0 \pi}{\beta}, \;\; \mbox{where}   \; \;  ({n_0} = \pm 1, \pm 3, \pm 5, ... )  .
 \end{equation} 

We impose the MIT boundary condition at the plates, which prevents the flow of fermion current out of the plates, as follows
\begin{equation}\label{s3b} 
\left(1 + i {\gamma ^ \mu } n_{\mu} ^j \right)  \Psi (x) | _{(z = {z_j})}  = 0,
\end{equation}
where $n_{\mu}^j$  is the unit vector perpendicular to the plate located at $z_j$. We consider the plates to be located at $ z =  - \frac{L}{2} $  and $z =   \frac{L}{2}$, and solve the free Dirac equation in three spatial dimensions, subject to the above boundary conditions. We obtain the following condition for the discrete spatial modes in $z$ direction:  
\begin{equation}\label{s1000}
f(k_{n_1}) = k_{n_1} \cos(k_{n_1} L) + m \sin(k_{n_1} L) = 0.
\end{equation}
Note that the modes for the massive case are irregular, {\it i.e.}, not equally spaced. However, for the massless case the modes are regular and given by
\begin{equation} \label{s1b}
k_{n_1} = \frac{{n_1} \pi}{2 L}   \;\;\;\;      ({n_1} = 1, 3, 5,....) .
\end{equation}

Using Eqs.\ (\ref{s3}, \ref{s3a}), the expression for the free energy becomes
\begin{eqnarray}\label{s04}
F_{\mbox{\scriptsize bounded}}(T,L) = - 2 T A \int \frac{d^2  K_T}{ {\left(2 \pi\right)}^2}  \sum\limits_{{n_0} =  - \infty }^{\infty '}  {\sum\limits_{n_1} \ln \left[ {\left( \frac{{n_0} \pi }{\beta } \right)}^2  + \omega _{{n_1} , {K_T}}^2 \right]}, 
\end{eqnarray}
where $\omega _{{n_1} , {K_T}} = \sqrt{ k_{n_1}^2 + K_T^2 + m^2}$, the prime on the summation denotes restriction of the sum to odd integers, and $A$ denotes the area of the plates. We shall refer to this expression for free energy as the first form. 
A very commonly used alternate expression for the first form, which has an embedded analytic continuation, is the following
\footnote{This is obtained by replacing the logarithm using $$\mbox{Log}(A)=-\lim\limits_{s \to 0}  \frac{\partial }{\partial s}  \left[\frac{1}{\Gamma(s)}\int_{0}^{\infty} dt e^{-t A} t^{s-1} \right].$$ We like to emphasize that the above replacement includes an analytic continuation. To trace it, we note that the expression for the logarithm is obtained by first using the identity $\mbox{Log}(A)=-\lim\limits_{s \to 0}  \frac{\partial }{\partial s}  A^{-s}$. Next, $A^{-s}$ has been replaced using the Euler integral representation of the Gamma function, {\it i.e.}, $\Gamma(s)=\int_0^{\infty} dt e^{-t} t^{s-1}$. This integral is finite for $s>0$, while it admits an analytic continuation for $s\leq0$. }
\begin{equation}\label{s4}
F _{\mbox{\scriptsize bounded}}(T,L)= 2 T A \int \frac{d^2 K_T}{\left(2 \pi \right)^2} \sum\limits_{n_0 =  - \infty }^{\infty '} \sum\limits_{n_1} \lim\limits_{s \to 0}   \frac{\partial }{\partial s}  \int_0^\infty  
		\frac{e^{ - t \left[ {\left( \frac{{n_0} \pi}{\beta} \right)}^2 + 
				\omega _{{n_1} , {K_T}}^2 \right]}}{ \Gamma (s)   t^{1 - s}}   dt .
\end{equation}
The integral over $t$ embodies the analytic continuation. This is important since in this paper we need to keep track of all analytic continuations. The result of this expression depends, in principle, on the order in which the sums and integrals are performed. However, if we do the integral over $t$ last we obtain the analytic continuation of this expression, which is certainly finite and unique, regardless of the order of other sums and integral. If we do the integral over $t$ at any step other than last and we are interested in the analytic of this expression, we are going to need a supplementary analytic continuation at the end. 

Integrating over the transverse momenta, we obtain
\begin{equation}\label{s004}
F _{\mbox{\scriptsize bounded}}(T,L)=\frac{ T A}{2 \pi}  \sum\limits_{n_0 =  - \infty }^{\infty '} \sum\limits_{n_1} \lim\limits_{s \to 0}   \frac{\partial }{\partial s}  \int_0^\infty  
		\frac{e^{ - t \left[ {\left( \frac{{n_0} \pi}{\beta} \right)}^2 + 
				\omega _{{n_1}}^2 \right]}}{ \Gamma (s)   t^{2 - s}}   dt  ,
\end{equation}
where $\omega_{n_1}=\sqrt{k_{n_1}^2 + m^2}$. If we now integrate over $t$, we obtain the form which is almost invariably used in the ZFA:
	\begin{equation}\label{SZeta}
		F _{\mbox{\scriptsize bounded}}(T,L)= \frac{TA}{2\pi} \sum\limits_{n_0 =  - \infty }^{\infty '} \sum\limits_{n_1 = 1}^{\infty '} \lim\limits_{s \to 0}   \frac{\partial }{\partial s}  \frac{1}{s-1}  \left[ \left( \frac{{n_0} \pi}{\beta} \right)^2 + 
		\omega _{{n_1}}^2 \right]^{1 - s} .
\end{equation}

One can perform the sum over the Matsubara frequencies in the original expression for the free energy given in Eq.~(\ref{s04}), using the Principle of the Argument theorem~\cite{r49Ahlf.}, to obtain the usual form in statistical mechanics~\cite{r31Bord2.}: 
\begin{equation}\label{s5}
F_{\mbox{\scriptsize bounded}} =   - 2 A \int  \frac{d^2  K_T}{{\left( 2 \pi \right)}^2}   \sum\limits_{n_1}  \left[  \omega _{{n_1} , {K_T}} + 
2 T \ln\left( 1 + e^{ - \beta \omega _{{n_1} , {K_T}}} \right) \right] ,
\end{equation}
which we shall refer to as the second form of the free energy. This expression and the original first form given by Eq.~(\ref{s04}), do not contain any embedded analytic continuation. One advantage of this form is that the contribution of the zero temperature part is separated from the thermal correction part.


\section {The Casimir Free Energy for a Massless Fermion Field}\label{FCasimir}

\indent 
In this section, we calculate the Casimir free energy, using its fundamental definition, for a free massless Dirac field between two parallel plates, separated by a distance $L$, with the MIT boundary conditions. 
In Sec.~\ref{massive}, we generalize to the massive case, and verify that, as expected, its massless limit coincide with the results of this section. As mentioned above, the fundamental definition of $F_{\mbox{\scriptsize Casimir}}$ is the difference between the free energy of the system in the presence of nonperturbative conditions or constraints, and the one with no constraints, both being at the {\em same} temperature $T$ and having the same volume. The nonperturbative conditions or constraints include boundary conditions, background fields such as solitons, and nontrivial space-time backgrounds. For cases where the constraints are in the form of non-trivial boundary conditions, the free cases can be defined as the cases in which the boundaries have been placed at spatial infinities. For the latter cases, the fundamental definition can be written as
\begin{equation}\label{s6}
F_{\mbox{\scriptsize Casimir}}(T,L) = F_{\mbox{\scriptsize bounded}}(T,L) - F_{\mbox{\scriptsize free}}(T,L) ,
\end{equation}
where the dependence of $F_{\mbox{\scriptsize free}}$ on $L$ simply denotes the restriction of the volume of space considered. We expect this dependence to be linear for simple extensive thermodynamic quantities such as $F_{\mbox{\scriptsize free}}(T,L)$.
	
To calculate the free energy for a massless fermion, we use the first form presented in  Eq.~(\ref{s004}), along with Eq.~(\ref{s1b}), and obtain: 
\begin{equation}\label{s7}
F _{\mbox{\scriptsize bounded}}(T,L)=\frac{TA}{2 \pi}  \sum\limits_{n_0 =  - \infty }^{\infty '} \sum\limits_{n_1 = 1}^{\infty '} \lim\limits_{s \to 0}   \frac{\partial }{\partial s}  \int_0^\infty  
\frac{e^{ - t \left[ {\left( \frac{{n_0} \pi}{\beta} \right)}^2 + 
\left( \frac{{n_1} \pi}{2 L}\right)^2  \right]}}{ \Gamma (s)  t^{2-s}}   dt  .
\end{equation}
The primes on the summations denote restrictions to odd integers ${n_0}$ and ${n_1}$. First, we express the sums in  symmetrized forms\footnote{The symmetrized forms we use for the sums over Matsubara and spatial modes are
\begin{eqnarray}\label{s7a}
\sum\limits_{{n_0} =  - \infty }^{\infty '}  e^{ - t  {\left(\frac{{n_0} \pi }{\beta } \right)}^2} =  
 \frac{1}{2} \left[ \sum\limits_{{n_0} = - \infty}^{\infty} e^{-t\frac{ \left(  2 n_0 + 1\right)^2  \pi^2}{\beta^2}} +\sum\limits_{{n_0} = - \infty}^{\infty} e^{-t \frac{ \left(  2 n_0 - 1 \right)^2  \pi^2}{\beta^2} }
\right] ,\nonumber\\
\sum\limits_{{n_1} =  1 }^{\infty '}  e^{ - t  {\left(\frac{{n_1} \pi }{ 2 L } \right)}^2}  =
\frac{1}{4} \left[ \sum\limits_{{n_1} = - \infty}^{\infty} e^{-t\frac{ {\left(  {n_1} + \frac{1}{2}\right)}^2  \pi^2}{L^2}} +\sum\limits_{{n_1} = - \infty}^{\infty} e^{-t \frac{{ \left(  {n_1} - \frac{1}{2} \right)}^2  \pi^2}{L^2} }
\right] .\nonumber
\end{eqnarray}} 
and then use the Poisson summation formula\footnote{The Poisson summation formula (see, for example,~\cite{r43Stein., r432Stein., r43Gasq., r432Gasq.})
 for a continuous and bounded function $f$ on $\mathbb{R}$ can be expressed as
\begin{equation}
\sum\limits_{{n} =  - \infty }^{\infty}  f(n)  =\sum\limits_{{m} =  - \infty }^{\infty} \int_{- \infty}^{\infty}  dx f(x) e^{- i 2 \pi m x}   \nonumber .
\end{equation}} 
for the sums over Matsubara frequencies and the spatial modes to obtain
\begin{equation}\label{s7bb}
 \sum\limits_{{n_0} =  -\infty }^{\infty '} e^{ - t  \left(\frac{{n_0} \pi }{\beta} \right)^2}  =
\frac{\beta}{2\sqrt{ \pi  t}} + \frac{\beta}{\sqrt{\pi  t}} \sum\limits_{{n_0} = 1}^\infty  {( - 1)}^{n_0}  
e^{ - \frac{{n_0}^2   \beta^2}{4 t} } ,
\end{equation}
\begin{equation}\label{s7b}
\sum\limits_{{n_1} =  1 }^{\infty '} e^{ - t  {\left(\frac{{n_1} \pi }{2 L} \right)}^2}  =
\frac{L}{2\sqrt{ \pi  t}} + \frac{ L}{\sqrt{\pi  t}} \sum\limits_{{n_1} = 1}^\infty  {( - 1)}^{n_1}  
e^{ - \frac{{n_1}^2   L^2}{t} }.
\end{equation} 
Next, we evaluate the integral over $t$ and  $\lim\limits_{s \to 0}   \partial / \partial s$ for all terms except for the one resulting from the multiplication of the first terms on the right hand sides of Eqs.~(\ref{s7bb}, \ref{s7b}), which contains a divergent integral, to obtain the free energy between the two plates. Below, we express the results in a form in which the zero temperature part is separated from the thermal correction part, as follows
\begin{eqnarray}\label{s8}
F_{\mbox{\scriptsize bounded}}(T,L) &=& F_{\mbox{\scriptsize bounded}}(0,L)+\Delta  F_{\mbox{\scriptsize bounded}}(T,L),  \mbox{ where} \nonumber\\
F_{\mbox{\scriptsize bounded}}(0,L) &=& \frac{A L}{8 \pi^2} \lim\limits_{s \to 0}   \frac{\partial }{\partial s} \frac{1}{\Gamma (s)} \int_0^\infty  \frac{dt}{  t^{3 - s}} - \frac{7 A \pi^2}{2880 L^3},  \mbox{ and }\nonumber\\
\Delta F_{\mbox{\scriptsize bounded}}(T,L) &=& - \frac{7 A L \pi^2}{180} T^4  + \frac{8 }{ \pi^2}  \sum\limits_{{n_0} = 1 }^{\infty} \sum\limits_{{n_1} = 1}^{\infty} \frac{A L T^4 (-1)^{n_0+n_1}}{\left[n_0^2 +\left( 2 n_1 T L\right)^2 \right]^2}.
\end{eqnarray}
Note that the first term of $\Delta F_{\mbox{\scriptsize bounded}}(T,L)$ is equivalent to the black-body radiation term.


On the other hand, the free energy of the unconstrained case, which is considered at the same temperature $T$ and same volume $V=AL$, can be computed using any of the forms presented in Sec.~\ref{Helmholtz free energy}. However, to ultimately use the fundamental definition, it is important to use the same form and same order of summations and integrations as used to calculate the free energy of the bounded configuration, which in this case is Eq.~(\ref{s004}). Hence, the expression that we use for free case is given by
\begin{eqnarray}\label{s9}
&&\hspace{-8mm}F_{\mbox{\scriptsize free}}(T,L)  = \frac{ T A}{2\pi}   \sum\limits_{{n_0} =  - \infty }^{\infty '} \int_0^\infty {\frac{L dk}{ \pi}} \lim\limits_{s \to 0} \frac{\partial }{\partial s}  \int_0^\infty 
\frac{e^{ - t  \left[\left(\frac{{n_0} \pi }{\beta } \right)^2 +  k^2 + m^2 \right]} }{ \Gamma (s) t^{2 - s}} dt ,
\end{eqnarray}
with $m=0$ in this case. Performing the same procedure as for the bounded case, we obtain the free energy of the free case. Below, we again express the results in a form in which the zero temperature part is separated from the thermal correction part, as follows
\begin{eqnarray}\label{s10aa}
F_{\mbox{\scriptsize free}}(T,L) &=& F_{\mbox{\scriptsize free}}(0,L) + \Delta F_{\mbox{\scriptsize free}}(T,L) , \mbox{ where} \nonumber \\
F_{\mbox{\scriptsize free}}(0,L)&=& \frac{A L}{8 \pi^2} \lim\limits_{s \to 0}   \frac{\partial }{\partial s} \frac{1}{\Gamma (s)} \int_0^\infty  \frac{dt}{ t^{3 - s}} , \mbox{ and}\nonumber \\
\Delta F_{\mbox{\scriptsize free}}(T,L) &=& - \frac{7 A L \pi^2}{180} T^4 . 
\end{eqnarray} 
Comparing this equation with Eq.~(\ref{s8}) for $F_{\mbox{\scriptsize bounded}}(T,L)$, we see that $F_{\mbox{\scriptsize free}}(0,L)$ is identical to the first term of $F_{\mbox{\scriptsize bounded}}(0,L)$, and $\Delta F_{\mbox{\scriptsize free}}(T,L)$ is identical to the first term of $\Delta F_{\mbox{\scriptsize bounded}}(T,L)$. The latter two are equivalent to the black-body radiation term. When calculating the Casimir free energy based on its fundamental definition, {\it i.e.}, Eq.~(\ref{s6}), all of these terms completely cancel
\footnote{If we start with Eq.~(\ref{SZeta}) for the massless case, {\it i.e.}, $\omega_{n_1}=k_{n_1}= n_1 \pi /(2 L)$, and use the Abel-Plana summation formula, the bounded and free cases both include  $F_{\mbox{\scriptsize free}}(0,L)=\frac{A L \pi}{4} \lim\limits_{s \to 0} \left[\frac{\partial}{\partial s}\frac{\pi^{1-s}}{s-1} \int_0^\infty dk k^{3-2s}\right]$, which is now explicitly divergent. However, this makes no difference in the fundamental approach since, upon subtraction, they again cancel each other yielding the same result given by Eq.~(\ref{s11}).}, yielding
\begin{eqnarray}\label{s11}
F_{\mbox{\scriptsize Casimir}}(T,L) = - \frac{7 A \pi^2}{2880 L^3}  + \frac{8 A L T^4}{ \pi^2}  \sum\limits_{{n_0} = 1 }^{\infty}
\sum\limits_{{n_1} = 1}^{\infty} \frac{(-1)^{n_0+n_1}}{\left[n_0^2 +\left( 2 n_1 T L\right)^2 \right]^2} .
\end{eqnarray} 
One can now compute the sum over ${n_0}$ to obtain,
\begin{eqnarray}\label{s12}
 F_{\mbox{\scriptsize Casimir}}(T,L) =  \frac{A T}{4 \pi L^2}  \sum\limits_{{n_1} = 1}^\infty  {( - 1)}^{n_1}\frac{ 1+  (2 \pi {n_1} T L)  \coth \left( 2 \pi {n_1}T L \right)}{n_1^3 \sinh \left( 2 \pi {n_1}T L \right)}.
\end{eqnarray} 
The zero temperature limits of Eqs.~(\ref{s11}, \ref{s12}) yield the following well known result $F_{\mbox{\scriptsize Casimir}}(0,L) =E_{\mbox{\scriptsize Casimir}}(0,L) =-7\pi^2 A/(2880L^3)$. 

One can also start with the second form of the free energy given by Eq.~(\ref{s5}) and use the Abel-Plana summation formula to obtain exactly the same expression for the Casimir free energy as given by Eq.~(\ref{s12}) (see Appendix \ref{appendixC:FCasimir}). 


In figure\ (\ref{fp5}), the Casimir free energy is plotted as a function of temperature for various values of $L$. As can be seen from this figure, $F_{\mbox{\scriptsize Casimir}}(T,L)$ is always negative, and goes to zero as the temperature or $L$ increases. Note that the vanishing of $F_{\mbox{\scriptsize Casimir}}(T,L)$ as $T$ goes to infinity occurs due to the subtraction of the free case at the same temperature, which amounts to the complete cancellation of the black-body term, without the need for any extra renormalization program. Moreover, this shows that there is no classical term proportional to $T$ for the massless fermions between plates, which, as we shall show, also holds for the massive fermions.
\begin{center}
	\begin{figure}[h!] 
		\includegraphics[width=13.5cm]{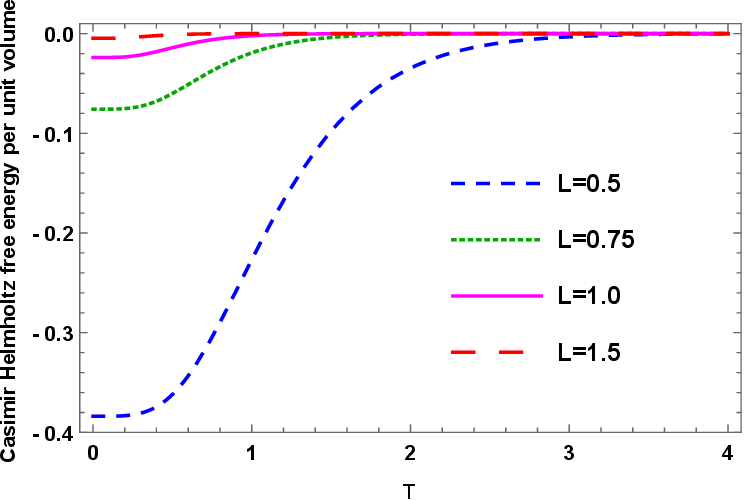}
		\caption{\label{fp5} \small
			The Casimir free energy per unit volume for a massless fermion field between two parallel plates as a function of temperature for various values of plate separations $L = \{ 0.5,0.75,1.0,1.5\} $.
	As an illustration of the magnitudes, for $L=1 \mu m$, each unit of the horizontal axis is equivalent to $458 K$, and each unit of the vertical axis is about $0.02 mPa$.}
	\end{figure}
\end{center}

\begin{center}
\begin{figure}[h!] 
\includegraphics[width=13.5cm]{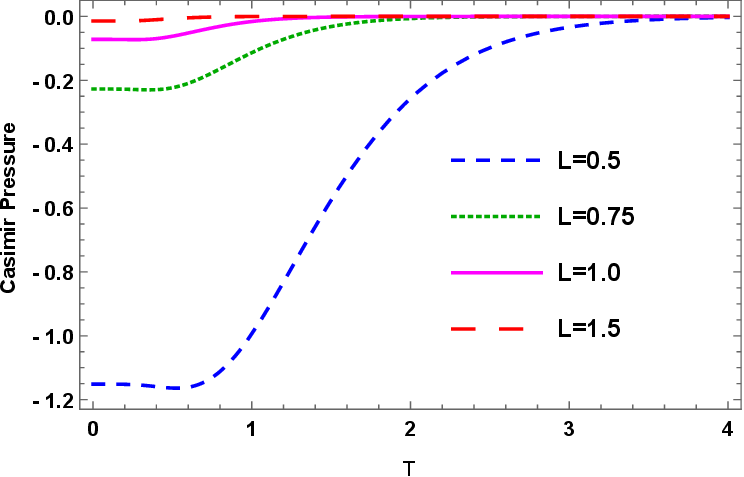}
\caption{\label{fp34} \small
The Casimir pressure for a massless fermion field between two parallel plates as a function of temperature for various values of plate separations $L = \{ 0.5,0.75,1.0,1.5\} $.
As an illustration of the magnitudes, for $L=1 \mu m$, each unit of the horizontal axis is equivalent to $458 K$, and each unit of the vertical axis is about $0.16 mPa$.}
\end{figure}
\end{center}

Having obtained the Casimir free energy, one can easily calculate all other thermodynamic quantities such as the Casimir pressure, energy, and entropy. For example the Casimir pressure is given by,
\begin{eqnarray}\label{s13}
P_{\mbox{\scriptsize Casimir}} (T,L) =  -\frac{1}{A}\frac{\partial}{\partial L} F_{\mbox{\scriptsize Casimir}}(T,L) = \frac{T}{2 \pi  L^3}  \sum\limits_{{n_1} = 1}^\infty  
	\frac{{( - 1)}^{n_1}}{n_1^3  \sinh  \left( 2 \pi {n_1}T L \right)}      \times  \nonumber\\
\bigg\{ 1 + (2 T L \pi {n_1}) \coth \left( 2 \pi {n_1}T L \right) +
	2 { \left( \pi {n_1}T L \right)}^2  
	\left[ 2 \coth^2  \left( 2 \pi {n_1}T L \right) - 1 \right] \bigg\} . \nonumber\\ 
\end{eqnarray}
Moreover, one can calculate directly the Casimir pressure based on its fundamental definition, as given by~\cite{r2Lif., r5Mehr.}, which is the differences between the pressure inside the two plates and outside the plates. To this end, we consider two inner plates enclosed within two outer plates, as the distance of the latter goes to infinity, and obtain the same result as given by Eq.~(\ref{s13}). By integrating over the distance between two plates, at fixed temperature, the Casimir free energy can be calculated, yielding the same result as given by Eq.~(\ref{s12}), without any extra terms. In figure\ \ref{fp34}, the Casimir pressure is plotted as a function of temperature for various values of $L$. As can be seen from this figure, $P_{\mbox{\scriptsize Casimir}}(T,L)$ is always negative, and goes to zero as the temperature or $L$ increases.  

\section {Massless Fermions and the Generalized Zeta Function}\label{zeta}
\indent
In this section, we consider the commonly used zeta function approach (ZFA) for computing the Casimir free energy for massless fermions at finite temperature. To fully explore this approach, we consider three different ways of using the zeta function and show that they yield equivalent results. Moreover, we shall also compute the Casimir free energy using the zero temperature subtraction approach (ZTSA)~\cite{r22Rav.}, and show that in the massless case its results are identical to those of the ZFA. However, as we shall show, the unrenormalized results are not equivalent to the one obtained in the last section, based on the fundamental definition of the Casimir free energy, and do not go to zero at high temperatures. We then illustrate how the renormalization procedure in this trivial example yields the correct results. Moreover, we show that if we calculate the free energies of the both the bounded and free cases using the zeta function and subtract them according to the fundamental approach, we again obtain the correct result.

For the computation of the Casimir free energy using the ZFA, we use only the first form given by Eq.~(\ref{s4}), as expressed in Eq.~(\ref{SZeta}). For the massless case we have $\omega_{n_1}=k_{n_1}= n_1 \pi /(2 L)$. As mentioned before, both sums in the free energy are over odd integers, which can be written as the difference between sums over all integers and even integers:
\begin{eqnarray}\label{s21}
&&F (T,L)= \frac{T A}{4 \pi }  \lim\limits_{s \to 0} \frac{\partial }{\partial s}  \frac{1}{s-1} \sum\limits_{n_{0} =  - \infty }^\infty  \sum\limits_{n_{1} =  - \infty }^{\infty } \left[ \left( \frac{n_0^2 \pi^2 }{\beta^2} + \frac{n_1^2 \pi^2 }{4 L^2 }\right)^{1 - s}  -   \right.   \nonumber\\
&& \hspace{-8mm}\left. \left( \frac{n_0^2 \pi^2 }{\beta^2} + \frac{n_1^2 \pi^2 }{ L^2 }\right)^{1 - s}   - \left( \frac{4 n_0^2 \pi^2 }{\beta^2} + \frac{n_1^2 \pi^2 }{4 L^2 }\right)^{1 - s} +  \left( \frac{4 n_0^2 \pi^2 }{\beta^2} + \frac{n_1^2 \pi^2 }{L^2 }\right)^{1 - s}   \right] .
\end{eqnarray}
To use the generalized zeta function, we have to impose the constraint that the double sums should not include the $({n_0}=0,{n_1}=0)$ mode. For cases in which the spatial modes do not include a zero mode, this constraint is automatically satisfied, otherwise this would amount to a renormalization. Our case is in the category of the former, as is apparent from the original expression Eq.~(\ref{s04}), due to antiperiodicty conditions on fermions and this shows up as the cancellation of the (0,0) modes between the four terms in Eq.~(\ref{s21}).

For our first method, we use the homogeneous generalized zeta function to do simultaneously the double summations for each of the four terms in Eq.~(\ref{s21}). In this case, the analytic continuation is rendered by the reflection formula (see Appendix \ref{appendixA:The zeta function}). Exactly the same method has been used in~\cite{r43RavFluc.} to obtain an expression for the Casimir free energy for a massless field confined between two plates. The final result of~\cite{r43RavFluc.} has been presented as a finite fractional expression which includes a double sum over $n_0$ and $n_1$. Here, we simplify these summations (see Appendix \ref{appendixA:The zeta function}), compute the sum over the Matsubara frequencies, and present the final result as follows 
\begin{eqnarray}\label{s23}
&&F_{\mbox{\scriptsize Zeta}}(T,L) = \Delta F_{\mbox{\scriptsize free}}(T,L) + \frac{A T}{16 \pi  L^2}  \sum\limits_{{n_1} = 1}^\infty   \frac{1}{n_1^3} 
\bigg\{  \csch \left(4 \pi T L {n_1} \right) - \nonumber\\ 
&& \hspace{-1cm}\left.   4  \csch \left(2 \pi T L {n_1} \right) + 4 T L \pi {n_1} \left[ \frac{\coth\left(4 \pi T L {n_1} \right)}{\sinh \left(4 \pi T L {n_1} \right)} - 2 \frac {\coth \left(2 \pi T L {n_1} \right)}{\sinh \left(2 \pi T L {n_1} \right)} \right] \right\rbrace,
\end{eqnarray}
where $\Delta F_{\mbox{\scriptsize free}}(T,L)$ is the thermal correction term of the massless free case, which is the black-body term proportional to $T^4$, given in Eq.~(\ref{s10aa}). The zero temperature limit of this expression gives the correct result for $F_{\mbox{\scriptsize  Casimir}}(0,L)=E_{\mbox{\scriptsize Casimir}}(0,L) = - 7\pi^2 A/(2880L^3)$. We have denoted the Casimir free energy obtained by this method as $F_{\mbox{\scriptsize Zeta}}$, to distinguish it from the one obtained using the fundamental definition, which we have simply denoted by $F_{\mbox{\scriptsize Casimir}}$.  

For our second method, we use the inhomogeneous form of the zeta function to sum over the spatial modes for each of the four terms in Eq.~(\ref{s21}), yielding (see Appendix \ref{appendixA:The zeta function})
\begin{eqnarray}\label{s21b}
&&F_{\mbox{\scriptsize Zeta}}(T,L) = -\frac{7 T A L}{2\sqrt{\pi^3} }  \lim\limits_{s \to 0} \frac{\partial }{\partial s} \frac{\Gamma\left(s - \frac{3}{2} \right)}{\Gamma (s)}  \sum\limits_{{n_0} =  1 }^\infty  \left( \frac{{n_0} \pi}{\beta} \right)^{3 - 2 s}  +\nonumber \\
 &&\frac{T A}{4 \pi {L^2}}  \sum\limits_{{n_1} = 1}^ \infty (-1)^{n_1} \frac{1 + \left( 2 \pi T L {n_1} \right) \coth (2 \pi T L {n_1})}{n_1^3  \sinh (2 \pi T L {n_1})}  .
\end{eqnarray}
The first term on the right hand side has a sum over temperature modes which is divergent, and the second one is identical to our result~(\ref{s12}). We can express this sum in terms of zeta function $\zeta(2s-3)$, the analytic continuation of which eventually leads to the following finite result\footnote{We have used: $\zeta(-3)=1/120$.}
\begin{equation}\label{s21bb}
F_{\mbox{\scriptsize Zeta}}(T,L) = \Delta F_{\mbox{\scriptsize free}}(T,L) + \frac{T A}{4 \pi {L^2}}  \sum\limits_{{n_1} = 1}^ \infty  (-1)^{n_1} \frac{1 + \left( 2 \pi T L {n_1} \right) \coth (2 \pi T L {n_1})}{n_1^3  \sinh (2 \pi T L {n_1})}.
\end{equation}
The first term is again the black-body term given by Eq.~(\ref{s10aa}). The zero temperature limit of this expression gives the correct result for $F_{\mbox{\scriptsize  Casimir}}(0,L)=E_{\mbox{\scriptsize Casimir}}(0,L) = - 7\pi^2 A/(2880L^3)$.

For our third method, we use the inhomogeneous form of the zeta function to sum over the Matsubara frequencies for each of the four terms in Eq.~(\ref{s21}), and obtain (see Appendix \ref{appendixA:The zeta function})
\begin{eqnarray}\label{s21c}
F_{\mbox{\scriptsize Zeta}}(T,L) &=& -\frac{7 A}{32\sqrt{\pi^3} }  \lim\limits_{s \to 0} \frac{\partial }{\partial s} \frac{\Gamma\left(s - \frac{3}{2} \right)}{\Gamma (s)}  \sum\limits_{{n_1} =  1 }^\infty  \left( \frac{{n_1} \pi}{L} \right)^{3 - 2 s}   +
 \nonumber \\
 && \frac{T^3 A}{\pi }  \sum\limits_{j = 1}^ \infty (-1)^j \frac{1 + \left(\frac{ \pi j }{2 T L} \right) \coth \left(\frac{ \pi j }{2 T L} \right)}{j^3  \sinh \left(\frac{ \pi j }{2 T L} \right)}.
\end{eqnarray}
If we again use the zeta function for the first sum as an analytic continuation, similarly to the previous case, we obtain,
\begin{equation}\label{s21d}
F_{\mbox{\scriptsize Zeta}}(T,L) = - \frac{7 A {\pi}^2}{ 2880  L^3 }  +
\frac{T^3 A}{\pi }  \sum\limits_{j = 1}^ \infty  (-1)^j \frac{1 + \left(\frac{ \pi j }{2 T L} \right) \coth \left(\frac{ \pi j }{2 T L} \right)}{j^3  \sinh \left(\frac{ \pi j }{2 T L} \right)}.
\end{equation} 
The first term is the zero temperature part, while the black-body $T^4$ term is embedded in the high temperature limit of the second term. It can be easily shown that the three expressions obtained by the three different methods of using the zeta function, {\it i.e.}, Eqs.\ (\ref{s23}, \ref{s21bb}, \ref{s21d}), are equivalent. While the last two methods have not been used before in the literature for obtaining the Casimir free energy, as far as we know, it is important to see that various ways of utilizing the zeta function yield equivalent results.

Next, we calculate the Casimir free energy using the zero temperature subtraction approach (ZTSA)~\cite{r22Rav.}. This approach is defined by
\begin{equation}\label{s16}
F_{\mbox{\scriptsize ZTSA}}(T,L) = F_{\mbox{\scriptsize bounded}} (T,L) - F_{\mbox{\scriptsize free}} (0,L)  .
\end{equation}
We present a method for calculating this quantity and outline four other methods, all yielding equivalent results. The first four methods are based on the first form of the free energy, given by Eq.~(\ref{s004}), and the fifth is based on the second form given by Eq.~(\ref{s5}). In the first method, we represent the sum over spatial modes in the symmetrized form used in Sec.~\ref{FCasimir}, and use the Poisson summation formula, evaluate the integral over $t$, take the limit $s\to0$, to obtain
\begin{eqnarray}\label{s16a}
F_{\mbox{\scriptsize ZTSA}}(T,L) &=&  \frac{16  A L {\pi}^2}{3} T^4 \left[ \sum\limits_{{n_0}=0}^\infty \left( {n_0}+ \frac{1}{2}\right)^3 - \int_{0}^{\infty} dk  k^3 \right]  +  \nonumber\\
 & & \frac{T A}{4 \pi {L^2}} \sum\limits_{{n_1} = 1}^ \infty (-1)^{n_1} \frac{1 + \left( 2 \pi T L {n_1} \right) \coth (2 \pi T L {n_1})}{n_1^3  \sinh (2 \pi T L {n_1})}   .
\end{eqnarray}
Using the Abel-Plana formula for the first part of Eq.~(\ref{s16a}), the final result is identical to $F_{\mbox{\scriptsize Zeta}}(T,L)$ given by Eq.~(\ref{s21bb}). In the second method, we represent the sum over the spatial modes as the difference between sums over all integers and even integers, as used in Eq.~(\ref{s21}), and again use Poisson summation formula. The final result is identical to $F_{\mbox{\scriptsize Zeta}}(T,L)$ given by Eq.~(\ref{s23}). The third method is similar to the first, except we perform the sum over the Matsubara frequencies using the Poisson summation formula given by Eq.~(\ref{s7bb}), and the final result is identical to $F_{\mbox{\scriptsize Zeta}}(T,L)$ given by Eq.~(\ref{s21d}). In the fourth method, we use directly the definition given in Eq.~(\ref{s16}), with its terms explicitly calculated in Sec.~\ref{FCasimir} and given by Eqs.~(\ref{s8}, \ref{s10aa}). The final result is identical to $F_{\mbox{\scriptsize Zeta}}(T,L)$ given by Eq.~(\ref{s21bb}). For the fifth method, we use the second form of the free energy given by Eq.~(\ref{s5}) and use the Abel-Plana formula, and the result is identical to $F_{\mbox{\scriptsize Zeta}}(T,L)$ given by Eq.~(\ref{s21d}). Exactly the same method has been used in~\cite{r22Rav.} and the Casimir free energy obtained in~\cite{r22Rav.} is equivalent to Eq.~(\ref{s21d}). They also obtained the Casimir free energy by calculating and using the Casimir pressure. Their final result is identical to Eq.~(\ref{s23}). 

\begin{center}
	\begin{figure}[h!]
		\includegraphics[width=13.5cm]{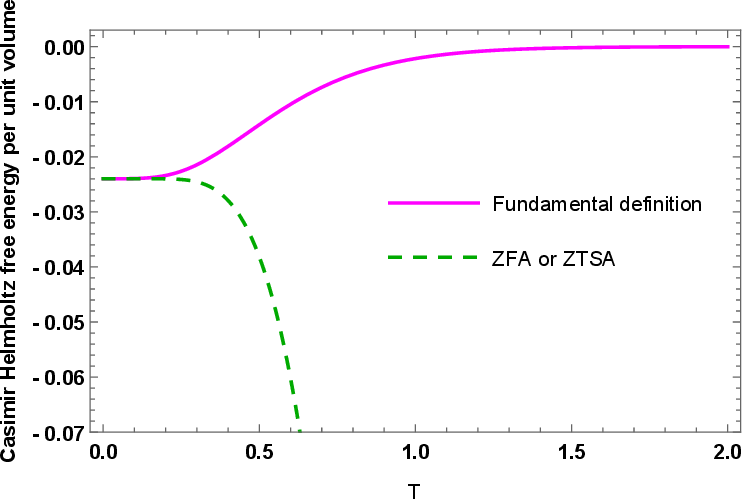}
		\caption{\label{fp6} \small
	The Casimir free energy per unit volume for a massless fermion field between two parallel plates as a function of temperature with fixed plate separation $L = 1.0$. The solid line is for the one obtained via the fundamental definition, and the dashed line is for the ones obtained by the zeta function approach ZFA, or the zero temperature subtraction approach ZTSA.}
	\end{figure}
\end{center}

So far, in this section, we have shown that for massless fermions, the results obtained using the ZFA are identical to that of the ZTSA. In other words, we have shown that for the massless cases at finite temperature, the use of the ZFA, which utilizes analytic continuation, is equivalent to the subtraction of the divergent zero temperature contribution to the free case. However, these results are not equivalent to the one obtained in the last section using the fundamental definition of the Casimir free energy which is given by Eq.~(\ref{s12}). We can summarize our results for the massless case as follows 
\begin{equation}\label{s19aa}
F_{\mbox{\scriptsize Zeta}}(T,L)=F_{\mbox{\scriptsize ZTSA}}(T,L)= F_{\mbox{\scriptsize  Casimir}}(T,L) + \Delta F_{\mbox{\scriptsize free}}(T,L),
\end{equation}
where two equivalent expressions for $F_{\mbox{\scriptsize  Casimir}}(T,L)$ are given by Eqs.(\ref{s12}, \ref{C4}). The difference is the thermal correction to the free energy of the free case, {\it i.e.}, $\Delta F_{\mbox{\scriptsize free}}(T,L) =- (7 A L \pi^2/180) T^4$, as given in Eq.~(\ref{s10aa}), which is equivalent to the black-body term. This difference can be traced back to the fact that the free energy of the free case at finite temperature contains the black-body term, the subtraction of which is included in the fundamental definition of the Casimir free energy, but it is not included in the ZFA or the ZTSA. We compare these results in Fig~(\ref{fp6}). As can be seen from this figure, the free energy obtained via the ZFA or ZTSA decreases as $T^4$, while the one obtained via the fundamental definition goes to zero at high temperatures. 

\begin{center}
	\begin{figure}[h!] 
		\includegraphics[width=13.5cm]{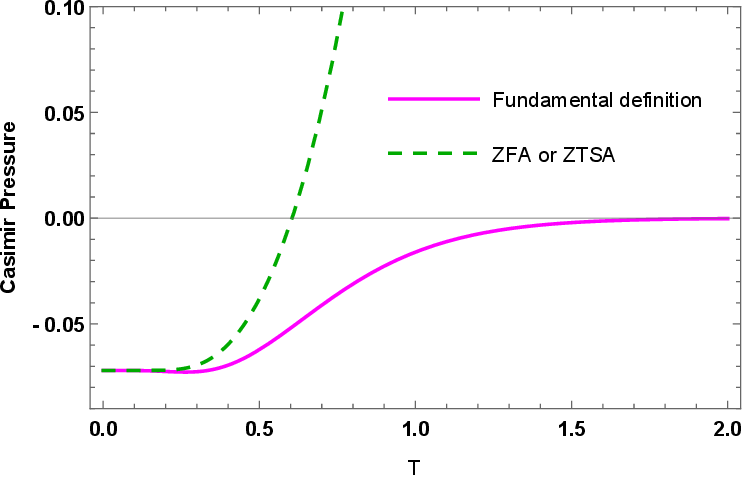}
				\caption{\label{fp7} \small
			The Casimir pressure for a massless fermion field between two parallel plates as a function of temperature with fixed plate separation $L = 1.0$. The solid line is for the pressure obtained via the fundamental definition, and the dashed line is for the pressure obtained by the zeta function approach ZFA, or the zero temperature subtraction approach ZTSA.}
	\end{figure}
\end{center}

One can now easily calculate all other thermodynamic quantities using the expressions obtained for the free energies by the ZFA or ZTSA. For example, calculation of pressure, using first part of Eq.~(\ref{s13}), yields
\begin{eqnarray}\label{s19}
P_{\mbox{\scriptsize Zeta}}(T,L) = P_{\mbox{\scriptsize ZTSA}}(T,L)=
	P_{\mbox{\scriptsize Casimir}}(T,L) +\Delta P_{\mbox{\scriptsize free}}(T),
\end{eqnarray}
where an expression for $P_{\mbox{\scriptsize  Casimir}}(T,L)$ is given by Eq.~(\ref{s13}), and $\Delta P_{\mbox{\scriptsize free}}(T) = (7 {\pi}^2/180) T^4$ is the thermal correction to the pressure of the free case. In figure~(\ref{fp7}), we compare the pressure obtained using the ZFA or the ZTSA, given by Eq.~(\ref{s19}), with the Casimir pressure obtained based on the fundamental definition given by Eq.~(\ref{s13}). As can be seen, the pressure obtained using the ZFA is negative, corresponding to attractive forces, at low temperatures and becomes positive, corresponding to repulsive forces, at high temperature, while the Casimir pressure is always negative and goes to zero at high temperatures. The difference between the two results~(\ref{s19}, \ref{s13}) is due to the pressure of the black-body term. 


As mentioned in the Introduction, it has been recognized that the ZFA might yield additional unphysical terms, and renormalization programs have been devised to subtract polynomials in $T$ appearing in the large temperature limit~\cite{r17Geyer., r29Junji.}. These are usually calculated using the heat kernel coefficients. In this case, the only nonzero term is the mononomial $T^4$ term, which is equivalent to the black-body term, the subtraction of which yields the correct results, based on the fundamental approach. Specifically, the removal of $\Delta F_{\mbox{\scriptsize free}}(T,L) = - (7 A L \pi^2/180) T^4$ from the expression for $F_{\mbox{\scriptsize Zeta}}(T,L)=F_{\mbox{\scriptsize ZTSA}}(T,L)$ in Eq.~(\ref{s19aa}), and the removal of $\Delta P_{\mbox{\scriptsize free}}(T) = (7 {\pi}^2/180) T^4$ from the expression for $P_{\mbox{\scriptsize Zeta}}(T,L)=P_{\mbox{\scriptsize ZTSA}}(T,L)$ in Eq.(\ref{s19}), yield the correct the results. We like to emphasize that these extra unphysical terms appear in the results of the ZFA and the ZTSA for different reasons. In the former case, they are left out by the embedded analytic continuation, and in the latter case, they are left out by its definition. As we have shown, in the massless case, these extra terms are simple polynomials in $T$ which can be easily removed by the renormalization programs that have been devised. 
In the next section we solve the massive case using the fundamental approach, which completely subtracts the corresponding thermodynamic quantities of the free case from those of the bounded case. We find that the thermal corrections to the free case are no longer simple polynomials that can be removed by any renormalization programs thus far devised to supplement ZFA or ZTSA.

\section {The Casimir Free Energy for a Massive Fermion Field}\label{massive}
\indent
In this section, we calculate the Casimir free energy, using its fundamental definition as given by Eq.~(\ref{s6}), for a massive fermion field confined between two parallel plates with the MIT boundary conditions at finite temperature. Then, we calculate other Casimir thermodynamic quantities, including pressure, energy, and entropy, and show that all of them are finite and go to zero as the temperature, mass, or $L$ increases. In the next section, we compute the Casimir free energy using the zeta function approach (ZFA) and also using the zero temperature subtraction approach (ZTSA)~\cite{r22Rav.}, and compare the results. 

We start with the first form of the free energy given by Eq.~(\ref{s004}). Then, we use the Poisson summation formula on the Matsubara frequencies\footnote{The form of the Poisson summation we have used is:
\begin{eqnarray}
\sum\limits_{{n_0} =  - \infty }^{\infty '}  e^{ - t (\frac{n_0 \pi }{\beta })^2}  = 
\frac{\beta}{\pi} \int_{0}^\infty  e^{ - t y^2} dy +  \frac{4}{\pi} \sum\limits_{{n_0} =  1 }^\infty \frac{(-1)^{n_0}}{n_0} \int_{0}^\infty y t  e^{ - t y^2} \sin(y {n_0} \beta) dy . \nonumber
\end{eqnarray}
}, evaluate the integral over $t$, and obtain
\begin{eqnarray}\label{s24} 
F_{\mbox{\scriptsize bounded}}(T,L) &=& \frac{A}{ \sqrt{\pi^3}} \sum\limits_{{n_1}>0} \lim\limits_{s \to 0} \frac{\partial}{\partial s}  \frac{1}{\Gamma (s)} \left[ \frac{\Gamma \left(s - \frac{3}{2}\right)}{4} \omega_{n_1}^{3 - 2s}	+\right.\nonumber \\
&& \left. \sum\limits_{{n_0} = 1}^\infty   (-1)^{n_0} \left(\frac{2\omega_{n_1}}{n_0 \beta}\right)^{\frac{3}{2} - s}
 K_{\frac{3}{2} - s} \left(\beta {n_0} \omega_{n_1}\right) \right] , 
\end{eqnarray}
where $\omega_{n_1}=\sqrt{{k_{n_1}}^2 + m^2}$.
 The spatial modes $k_{n_1}$ are the roots of $f(k_{n_1})$ in Eq.~(\ref{s1000}), which for the massive case are irregular, {\it i.e.}, they are not equally spaced. To evaluate the sum over the spatial modes, we use the Principle of the Argument theorem and after simplifying (see Appendix \ref{appendixB:The Argument Principle}), we can express the free energy of the bounded region as
\begin{eqnarray}\label{s1003}
&&\hspace{-8mm}F_{\mbox{\scriptsize bounded}}(T,L) =  \frac{A }{\sqrt{\pi^5}} \lim\limits_{s \to 0} \frac{\partial}{\partial s} \left\lbrace  L \frac{\Gamma \left(s - \frac{1}{2}\right) }{2 \Gamma (s)} \int_{ 0}^{ \infty}p^{2 - 2s} \omega(p) dp	-\sum\limits_{n_0 = 1}^{\infty} (-1)^{n_0} \times\right.\nonumber \\
&&\hspace{-10mm}\left.  \left(\frac{2}{n_0 \beta}\right)^{\frac{1}{2} - s}\int_{0}^{\infty} \frac{ L p^{\frac{3}{2} - s}}{\Gamma (s)} J_{\frac{1}{2} - s}  \left(n_0 \beta p\right) \omega(p) dp+ \int_{0}^{\infty} \ln\left( 1 + \frac{\omega(p)  -   m}{\omega(p)  + m} e^{ - 2 L \omega(p)}\right) \times\right.\nonumber \\
&&\left.   \left[\frac{\Gamma \left(s - \frac{1}{2}\right) }{2 \Gamma(s)}p^{2 - 2s} - \sum\limits_{n_0 = 1}^{\infty} (-1)^{n_0} \left(\frac{2}{n_0 \beta}\right)^{\frac{1}{2} - s} \frac{ p^{\frac{3}{2} - s} J_{\frac{1}{2} - s}  \left(n_0 \beta p\right)}{\Gamma (s)} \right]   dp \right \rbrace ,
\end{eqnarray} 
where $\omega(p)=\sqrt{p^2 + m^2}$. Only the first term of the above expression contains a divergent integral. Therefore, for the other terms, which include the logarithm function and the Bessel function, we take the derivative with respect $s$, take the limit $s \to 0$, and after simplifying we obtain
\begin{eqnarray}\label{s1003bv}
&&F_{\mbox{\scriptsize bounded}}(T,L) =  \frac{A L}{2\sqrt{\pi^5}} \lim\limits_{s \to 0} \frac{\partial}{\partial s}\frac{ \Gamma \left(s - \frac{1}{2}\right)}{\Gamma (s)} \int_{ 0}^{ \infty}  p^{2 - 2s}  \omega(p) dp	+ \frac{2A L T^2 m^2}{\pi^2} \times\nonumber \\
&& \sum\limits_{n_0 = 1}^{\infty} \frac{(-1)^{n_0}}{n_0^2} K_2 \left(n_0 \beta m\right) - \frac{A}{\pi^2} \int_{0}^{\infty} \ln\left( 1 + \frac{\omega(p)  -   m}{\omega(p)  + m} e^{ - 2 L \omega(p)}\right) \times \nonumber \\
&&  \left[ p^2 +2T \sum\limits_{n_0 = 1}^{\infty} \frac{(-1)^{n_0}}{n_0} p \sin \left(n_0 \beta p\right)\right]   dp  .
\end{eqnarray} 
Since we are going to use the fundamental definition of the Casimir free energy, we also need to calculate the free energy of the free massive case at finite temperature. We start with the first form of the free energy as given by Eq.~(\ref{s4}), use the Poisson summation on the Matsubara frequencies, and evaluate the integral over $t$. Then, we can express the free energy of the free case as a zero temperature part and a finite temperature correction part as follows 
\begin{eqnarray}\label{s25}
F_{\mbox{\scriptsize free}}(T,L) &=& F_{\mbox{\scriptsize free}}(0,L) +\Delta F_{\mbox{\scriptsize free}}(T,L) \mbox{, where} \nonumber\\
F_{\mbox{\scriptsize free}}(0,L) &=& \frac{A L}{2\sqrt{\pi^5}}\lim\limits_{s \to 0} \frac{\partial}{\partial s}\frac{ \Gamma \left(s - \frac{1}{2}\right)}{\Gamma (s)} \int_{ 0}^{ \infty}  k^2  \left[\omega(k)\right]^{1 - 2s} dk \mbox{, and} \nonumber\\
\Delta F_{\mbox{\scriptsize free}}(T,L) &=& \frac{2 A L T^2 m^2}{\pi^2} \sum\limits_{j = 1}^{\infty} \frac{(-1)^j}{j^2}  K_{2}\left( j \beta m\right) .
\end{eqnarray}
The first two terms of $F_{\mbox{\scriptsize bounded}}(T,L)$, given by Eq.~(\ref{s1003bv}), are equivalent to the two terms of $F_{\mbox{\scriptsize free}}(T,L)$, given by Eq.~(\ref{s25}). The second terms are actually identical, while the first terms contain equivalent divergent integrals. However the first terms, both being equal to $F_{\mbox{\scriptsize free}}(0,L)$, turn out to be finite with the method presented in this section\footnote{The equivalent divergent integrals in the first terms of Eqs.~(\ref{s1003bv}, \ref{s25}) can be computed using the dimensional regularization, with fixed $s$, yielding $ [- m^{4-2s}/(4\sqrt{\pi})] \Gamma(3/2-s)  \Gamma(s-2)$.
We can now compute these two terms exactly, obtaining $[A L m^4/(32 \pi^2)]  \left[ 3 - 4 \ln(mL)\right]$. This result is finite due to the analytic continuation embedded in expression that we have used for the first form of the free energy Eq.~(\ref{s4}), as mentioned in Sec.~\ref{Helmholtz free energy}.}.
Now, using the fundamental definition, as expressed in Eq.~(\ref{s6}), these four terms cancel each other upon subtraction, and the Casimir free energy for a massive fermion field confined between two plates becomes\footnote{The sum in Eq.~(\ref{s27}) can be written in closed form upon using:	
$\sum\limits_{j = 1}^{\infty} \frac{(-1)^j}{j}  \sin\left( j \beta p \right) = - {\tan}^{-1} \left[\tan\left(\frac{p \beta}{2}\right)\right]$} 
\begin{eqnarray}\label{s27}
F_{\mbox{\scriptsize Casimir}}(T,L) &=& - \frac{A}{\pi^2}\int_{0}^{\infty}  \left\{ p^2 + 2 T  p \sum\limits_{j = 1}^{\infty} \frac{(-1)^j}{j}  \sin\left( j \beta p \right)  \right\} \times   
\nonumber \\
 & & \ln\left( 1 + \frac{\omega(p) - m}{\omega(p)  + m}   e^{ - 2 L \omega(p)}\right)    dp    .
\end{eqnarray}
The zero temperature and finite temperature correction parts, {\it i.e.}, $F_{\mbox{\scriptsize Casimir}}(0,L)$ and $\Delta F_{\mbox{\scriptsize Casimir}}(T,L)$, are associated with the two terms in the curly bracket in Eq.~(\ref{s27}), respectively.
This expression is our main result for Casimir free energy which is obtained by its fundamental definition and, as far as we know, has not been presented heretofore.
 
One can easily show that using the second form of the free energy given by Eq.~(\ref{s5}), one obtains exactly the same expression as in Eq.~(\ref{s27}). We show the details of this computation, in which we utilize dimensional regularization, in Appendix \ref{appendixD:The Dimensional Regularization}. 
As we have shown, the free energies of both the bounded and free cases contain $F_{\mbox{\scriptsize free}}(0,L)$ which is in principle divergent. Its value, obtained using the second form in Appendix \ref{appendixD:The Dimensional Regularization}, is proportional to $\Gamma(-2)$ which is divergent. This is due to the fact that the second form given by Eq.~(\ref{s5}), similar to the expression of the first form given by Eq.~(\ref{s04}) and in contrast to the expression of the first form given by Eq.~(\ref{s4}), does not have an embedded analytic continuation. 
One of the advantages the fundamental definition is that $F_{\mbox{\scriptsize free}}(0,L)$ terms cancel upon subtraction of $F_{\mbox{\scriptsize free}}(T,L)$ from $F_{\mbox{\scriptsize bounded}}(T,L)$,  whether they are infinite or have been rendered finite by analytic continuations. The other advantage is that both $F_{\mbox{\scriptsize bounded}}(T,L)$ and $F_{\mbox{\scriptsize free}}(T,L)$ contain $\Delta F_{\mbox{\scriptsize free}}(T,L)$ terms which also cancel upon subtraction, regardless of whether it is a simple polynomial of $T$ or not.

When using the fundamental approach, we have implicitly assumed that the contributions to the Casimir free energy coming from the regions outside of the bounded region cancel with the corresponding contributions of the free case. We use the Boyer method to ascertain this cancellation in Appendix \ref{appendixE:The Boyer method}.

In figure~(\ref{fp8}), we plot the Casimir free energy of a massive fermion field for various values of mass. As can be seen, the Casimir free energy goes to zero rapidly as the temperature or mass of the fermion field increases. As can be seen directly from Eq.~(\ref{s27}), the Casimir free energy goes to zero rapidly as $L$ increases, as well. Moreover, as can be seen from figure~(\ref{fp8}), and can be shown easily from Eq.~(\ref{s27}), the massless limit of our result for the massive case coincides exactly with the massless case given by Eq.~(\ref{s12}). The zero temperature limit of $F_{\mbox{\scriptsize Casimir}}(T,L)$, given by Eq.~(\ref{s27}), yields the following well known result, as reported in, for example,~\cite{r44Mavaev., r44EliSanT.},
\begin{eqnarray}\label{s28}
\hspace{-9mm}F_{\mbox{\scriptsize Casimir}}(0,L) =E_{\mbox{\scriptsize Casimir}}(0,L) = - \frac{A}{\pi^2}\int_{0}^{\infty}   p^2  \ln\left[1 + \frac{\omega(p) - m}{\omega(p)  + m}   e^{ - 2 L \omega(p)}\right]    dp    .
\end{eqnarray} 
\begin{center}
\begin{figure}[h!] 
\includegraphics[width=13.5cm]{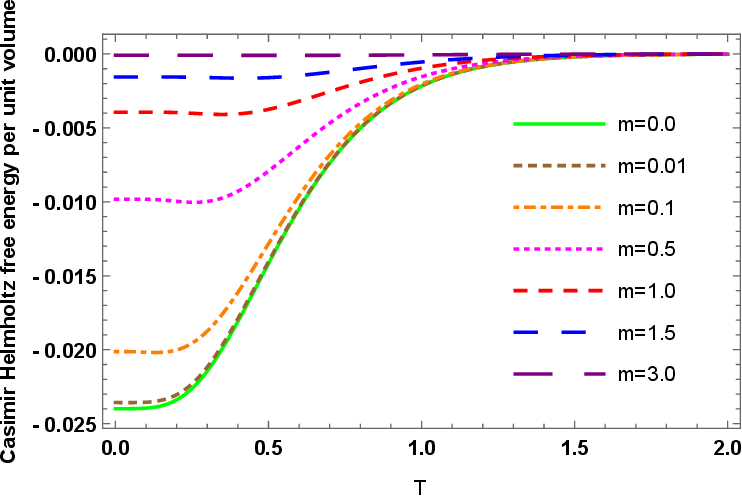}
\caption{\label{fp8} \small
The Casimir free energy per unit volume for a massive fermion field between two parallel plates as a function of temperature with fixed plate separation $L = 1.0$, for various values of mass $m = \{ 0.0,0.01,0.1,0.5,1.0,1.5,3.0 \} $. Note that the Casimir free energy goes to zero as the temperature or mass increases.}
\end{figure}
\end{center}

\begin{center}
\begin{figure}[h!] 
\includegraphics[width=13.5cm]{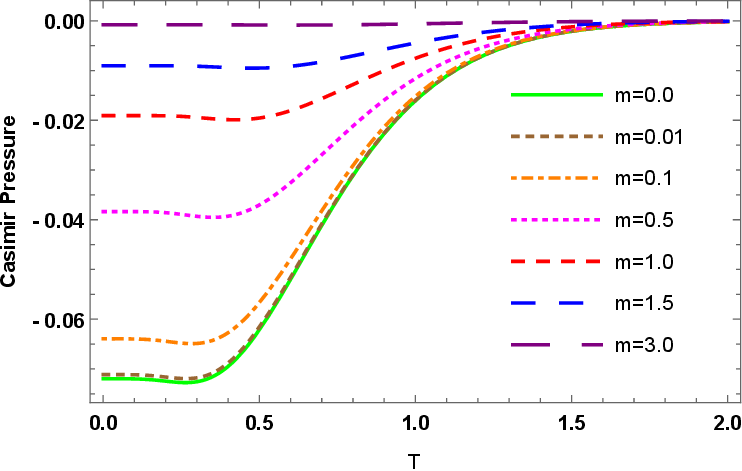}
\caption{\label{fp9} \small
The Casimir pressure for a massive fermion field between two parallel plates as a function of temperature with fixed plate separation $L = 1.0$, for various values of mass $m = \{ 0.0,0.01,0.1,0.5,1.0,1.5,3.0 \} $. Note that the Casimir pressure goes to zero as the temperature or mass increases.}
\end{figure}
\end{center}

Now, one can obtain other thermodynamic quantities including, the Casimir pressure, Casimir energy, and Casimir entropy from the expression we have obtained for the Casimir free energy in Eq.~(\ref{s27}). We calculate the Casimir pressure for a massive fermion field, in analogy with the massless case shown in Eq.~(\ref{s13}), and obtain,  
\begin{eqnarray}\label{s29}
P_{\mbox{\scriptsize Casimir}}(T,L) &=&  - \frac{1}{\pi^2} \int_{0}^{\infty} \left[\omega(p)^2 - m \omega(p)\right]  \left[ 1 - \tanh\left( L \omega(p)\right) \right] \times \nonumber\\
& & 
 \frac{ p^2 + (2 T  p) \sum\limits_{j = 1}^{\infty} \frac{(-1)^j}{j} \sin\left( j \beta p \right) } {\omega(p) + m \tanh\left( L \omega(p)\right)}  dp   .
\end{eqnarray}
The zero temperature and finite temperature correction parts, {\it i.e.}, $P_{\mbox{\scriptsize Casimir}}(0,L)$ and $\Delta P_{\mbox{\scriptsize Casimir}}(T,L)$, are associated with the two terms in the numerator of the fraction term in Eq.~(\ref{s29}), respectively. We plot $P_{\mbox{\scriptsize Casimir}}(T,L)$ for various values of mass in figure~(\ref{fp9}). As can be seen, the Casimir pressure also goes to zero rapidly as the temperature or the mass of fermion field increases. Moreover, as can be seen from figure~(\ref{fp9}), and can be shown easily from Eq.~(\ref{s29}), the massless limit of our result for the massive case coincides exactly with the massless case given by Eq.~(\ref{s13}).

\begin{center}
\begin{figure}[h!] 
\includegraphics[width=13.5cm]{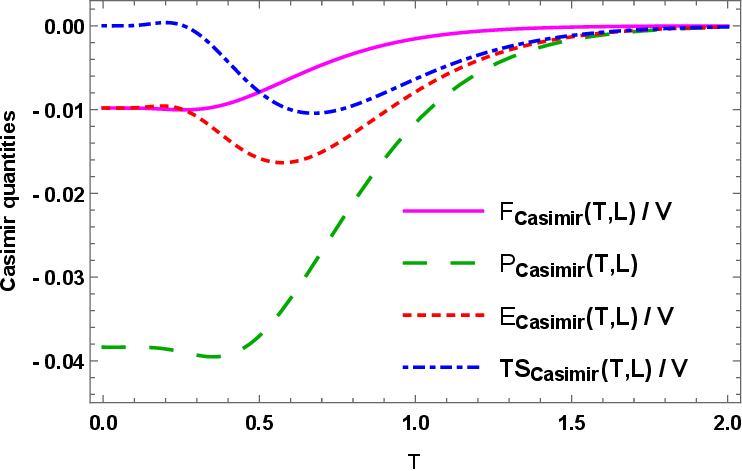}
\caption{\label{fp60} \small
The Casimir thermodynamic quantities, including the free energy, pressure, energy, and entropy, obtained using the fundamental approach, for a massive fermion field between two parallel plates as a function of temperature, with fixed plate separation $L = 1.0$, and mass $m = 0.5$. Note that all of the Casimir quantities go to zero as the temperature increases.}
\end{figure}
\end{center}

The Casimir energy can be calculated using either of the following two expressions,
\begin{equation}
E_{\mbox{\scriptsize Casimir}}(T,L) = E_{\mbox{\scriptsize bounded}}(T,L) - 
E_{\mbox{\scriptsize free}}(T,L)=\frac{\partial}{\partial \beta} \left[\beta F_{\mbox{\scriptsize Casimir}}(T,L)\right].
\end{equation}
The first expression is its fundamental definition. We use the second expression to obtain,
\begin{eqnarray}\label{s2009}
&& E_{\mbox{\scriptsize Casimir}}(T,L)= - \frac{A}{3 \pi^2}  \int_{0}^{\infty}  dp \left( L p^2 - m\right)  \left[ 1 - \tanh \left( L \omega(p)\right) \right] \times  \nonumber \\ 
&& \hspace{-1cm} \frac{  p^4 + 6 T^3  \sum\limits_{j = 1}^{\infty} \frac{(-1)^j}{j^3} p \Bigg\{  (2 j \beta p) \cos (j \beta p) + \left[ {(j \beta p)^2} - 2 \right] \sin (j \beta p)\Bigg\}  }{\omega(p)  \left( \omega(p) + m\right)  \left[ \omega(p) +  m  \tanh \left( L \omega(p)\right) \right] }       . 
\end{eqnarray}  
Finally, we calculate the Casimir entropy and obtain,
\begin{eqnarray}\label{s2020}
\hspace{-9mm}S_{\mbox{\scriptsize Casimir}}(T,L) =-\frac{\partial}{\partial T}F_{\mbox{\scriptsize Casimir}}(T,L)=  - \frac{2 A T^2}{\pi^2}  \int_{0}^{\infty}  dp  \left( L p^2 - m\right)   \times  \nonumber \\ 
\hspace{-9mm}\left[ 1 - \tanh \left( L \omega(p)\right) \right]\sum\limits_{j = 1}^{\infty} \frac{(-1)^j p \left[  (3 j \beta p) \cos (j \beta p) + [(j \beta p)^2 - 3] \sin (j \beta p)\right] }{ j^3\omega(p)  \left( \omega(p) + m\right)  \left[ \omega(p) +  m  \tanh \left( L \omega(p)\right) \right] }.  
\end{eqnarray}  
In figure~(\ref{fp60}), we show all of these Casimir thermodynamic quantities. Note that all of these quantities are finite and go to zero at high temperatures. In analogy with the case of Casimir free energy, one can easily show that all of the Casimir thermodynamic quantities also go to zero as $m$ or $L$ increases.
It is worth mentioning that the Casimir entropy in our model is negative for almost the entire range of $T$. Negative Casimir entropy has been reported earlier in Ref.~\cite{r45Bor.} for thin sheets, using the ZFA with the heat kernel method. In the fundamental approach used here, the Casimir entropy is $S_{\mbox{\scriptsize Casimir}}(T,L)= S_{\mbox{\scriptsize bounded}}(T,L)-S_{\mbox{\scriptsize free}}(T,L)$. Hence the interpretation of $S_{\mbox{\scriptsize Casimir}}(T,L)<0$ for almost the entire range of $T$ is simply that $S_{\mbox{\scriptsize bounded}}(T,L)< S_{\mbox{\scriptsize free}}(T,L)$ in that range, while both of them are positive here.

\section{Massive Fermions and the Generalized Zeta Function}\label{Epstein}

\indent
The zeta function approach (ZFA) has been used to calculate the Casimir free energy for the massive fermion field between two plates and some solutions have been presented (see for example in~\cite{r45Teo., r283Flach.}). In this section, we compute explicitly the final results for the Casimir free energy and Casimir pressure for this problem using the ZFA, and also using the zero temperature subtraction approach (ZTSA), and show that, contrary to the massless case, they yield different results. Most importantly, we show that neither of these results are equivalent to the one obtained in Sec. \ref{massive} based on the fundamental approach. Moreover, we show that these discrepancies cannot be fixed completely by the renormalization program mentioned before (see e.g.~\cite{r17Geyer., r29Junji.}), since the extra unphysical terms are non-polynomial functions of $T$.

We use the first form of the free energy given by Eq.~(\ref{SZeta}), and present the sum over Matsubara frequencies as the difference between sum over all integers and even integers, as used in Eq.~(\ref{s21}). Then, we use the inhomogeneous zeta function on the Matsubara frequencies (see Appendix \ref{appendixA:The zeta function}). The result is
\begin{eqnarray}\label{s33}
&&F_{\mbox{\scriptsize Zeta}}(T,L) = \frac{A}{4 \sqrt{\pi ^3}} \sum\limits_{{n_1}>0} \lim\limits_{s \to 0}  \frac{\partial }{\partial s} \frac{1 }{\Gamma(s)} \left\lbrace \Gamma \left(s - \frac{3}{2}\right) \left(\omega_{n_1}^2 \right)^{\frac{3}{2} - s}	+ \right. \nonumber \\
&& \left.   8\sum\limits_{{n_0} = 1}^\infty     \left(\frac{\omega_{n_1}}{\beta {n_0}} \right)^{\frac{3}{2} - s}\left[ K_{\frac{3}{2} - s}\left( 2 \beta {n_0} \omega_{n_1} \right) -  2^{\frac{1}{2} - s}
K_{\frac{3}{2} - s}\left(\beta {n_0} \omega_{n_1}\right) \right] \right\rbrace  .
\end{eqnarray}
To explore the mechanism of removal of divergences from this point forward, it is useful to compare this expression with the analogous one that we have obtained for the massless case after using the inhomogeneous zeta function on the Matsubara frequencies, {\it i.e.}, Eq.~(\ref{s21c}). The first term in both expressions includes a divergent sum over the spatial modes and is a leftover from the use of inhomogeneous zeta function on the Matsubara frequencies. In the massless case, the spatial modes were regular and we could obtain the analytic continuation of its divergent term using a supplementary zeta function. In the present case, the modes are irregular and, as before, we compute the sum over the spatial modes using the Principle of the Argument theorem (see Appendix \ref{appendixB:The Argument Principle}), and obtain the following expression
\begin{eqnarray}\label{s34}
&&\hspace{-8mm}F_{\mbox{\scriptsize Zeta}}(T,L) =  \frac{A }{2 \sqrt{\pi ^5}} \lim\limits_{s \to 0}  \frac{\partial }{\partial s} \frac{1}{\Gamma(s)}    \left\lbrace \int_{ 0}^{ \infty} dp \left[  L \omega(p)  +\ln\left( 1 + \frac{\omega(p) - m}{\omega(p)  + m}  e^{ - 2 L \omega(p)}\right)\right] \right. \nonumber\\ 
&& \hspace{-8mm} \left. \left[  \Gamma \left(s - \frac{1}{2}\right) p^{2 - 2s} -   \sum\limits_{{n_0} = 1}^\infty \frac{4 \pi p^{\frac{3}{2} - s}}{\left(n_0 \beta \right)^{\frac{1}{2} - s}}  \left[ 2 J_{\frac{1}{2} - s}\left(2 \beta {n_0} p\right) - 2^{\frac{1}{2} - s} J_{\frac{1}{2} - s}\left( \beta {n_0} p\right) \right]  \right] \right\rbrace  ,
\end{eqnarray}
where $\omega(p)=\sqrt{p^2 + m^2}$. The terms which include the logarithm function, are finite in the domain of integration. So, for these terms, we take the derivative with respect to $s$, take the limit $s \to 0$, and obtain a result which is identical to $F_{\mbox{\scriptsize Casimir}}$ as given by Eq.~(\ref{s27}), 
\begin{eqnarray}\label{s34b}
&&\hspace{-8mm} F_{\mbox{\scriptsize Zeta}}(T,L) = F_{\mbox{\scriptsize Casimir}}(T,L) +  \frac{AL }{2 \sqrt{\pi ^5}} \lim\limits_{s \to 0}  \frac{\partial }{\partial s} \frac{\Gamma \left(s - \frac{1}{2}\right)}{\Gamma(s)}    \int_{0}^{  \infty} dp \omega(p) \left\lbrace  p^{2 - 2s}  -\right.\nonumber \\
&&\hspace{-8mm}  \left. 4 \pi \sum\limits_{{n_0} = 1}^\infty \frac{ p^{\frac{3}{2} - s}}{\left(n_0 \beta \right)^{\frac{1}{2} - s}}   \left[ 2 J_{\frac{1}{2} - s}\left(2 \beta {n_0} p\right)- \sqrt{2^{\left(1- 2 s\right)}} J_{\frac{1}{2} - s}\left( \beta {n_0} p\right) \right]   \right\rbrace. 
 \end{eqnarray}
 The remaining terms are extra unphysical terms which are leftovers from the use of the inhomogeneous zeta function. The first of these terms includes a divergent integral and is identical to the first term of the $F_{\mbox{\scriptsize bounded}}(T,L)$, given by Eq.~(\ref{s1003bv}). As shown in Sec.~\ref{massive}, these terms are both equal to the analytic continuation of $F_{\mbox{\scriptsize free}}(0,L)$ which is equal to $[A L m^4/(32 \pi^2)]  \left[ 3 - 4 \ln(mL)\right]$. 
This is the analytic continuation embedded in the first form of free energy Eq.~(\ref{s4}), and has rendered this term finite.
Finally, taking the derivative with respect to $s$, and taking the limit  $s \to 0$, we can express the final result as follows 
\begin{equation}\label{s35d}
F_{\mbox{\scriptsize Zeta}}(T,L) = F_{\mbox{\scriptsize Casimir}}(T,L) + \frac{A L m^4}{32 \pi^2}  \left[ 3 - 4 \ln(mL) \right]+ \Delta F_{\mbox{\scriptsize free}}(T,L),
\end{equation}
where the explicit form of $F_{\mbox{\scriptsize Casimir}}(T,L)$, obtained by the fundamental definition, is given by Eq.~(\ref{s27}). The order of the terms presented above is the same as in Eq.~(\ref{s34b}), {\it i.e.}, the second term is the finite temperature-independent term mentioned above, and the last term is the thermal correction to the free energy of the free massive case, {\it i.e.}, $\Delta F_{\mbox{\scriptsize free}}(T,L)$ given in Eq.~(\ref{s25}).
This expression is our main result for Casimir free energy obtained by analytic continuation and, as far as we know, has not been presented heretofore.

Next, we calculate the Casimir free energy using the ZTSA~\cite{r22Rav.}, as defined in Eq.~(\ref{s16}). We use the results presented in Sec.\ \ref{massive} for the free energy of the bounded and free cases, given by Eqs.~(\ref{s1003}, \ref{s25}), to obtain 
\begin{eqnarray}\label{s31}
F_{\mbox{\scriptsize ZTSA}}(T,L)&=&F_{\mbox{\scriptsize Casimir}}(T,L) + \Delta F_{\mbox{\scriptsize free}}(T,L).
\end{eqnarray}
It is worth mentioning that the massless limit of $F_{\mbox{\scriptsize Zeta}}(T,L)$ and $F_{\mbox{\scriptsize ZTSA}}(T,L)$ for the massive cases, given by Eqs.~(\ref{s35d}, \ref{s31}), coincide exactly with their massless cases, given by Eq.~(\ref{s21bb}). 

Now we can compare the three different results obtained for the Casimir free energy using the fundamental definition, {\it i.e.,} $F_{\mbox{\scriptsize Casimir}}$ given by Eq.~(\ref{s27}), the zeta function approach, {\it i.e.,} $F_{\mbox{\scriptsize Zeta}}$ given by Eq.~(\ref{s35d}), and the zero temperature subtraction approach, {\it i.e.,} $F_{\mbox{\scriptsize ZTSA}}$ given in Eq.~(\ref{s31}). Comparing $F_{\mbox{\scriptsize Zeta}}$ with $F_{\mbox{\scriptsize ZTSA}}$, we observe that, contrary to the massless case, the results are not equivalent: there is an extra term in $F_{\mbox{\scriptsize Zeta}}$, which is its second term in Eq.~(\ref{s35d}) and is temperature-independent. As mentioned above, this extra term is an analytic continuation of the divergent term which appears after using the inhomogeneous Epstein zeta function. Next, we compare these two results with $F_{\mbox{\scriptsize Casimir}}$. First, as can be seen from the Eqs.~(\ref{s27}, \ref{s35d}, \ref{s31}), $F_{\mbox{\scriptsize Casimir}}$ does not include the extra temperature independent term in $F_{\mbox{\scriptsize Zeta}}$, mentioned above. Second,  $F_{\mbox{\scriptsize Casimir}}$ does not include the thermal correction term of the free case, {\it i.e.}, $\Delta F_{\mbox{\scriptsize free}}(T,L)$ given by Eq.~(\ref{s25}), which appears in both $F_{\mbox{\scriptsize Zeta}}$ and $F_{\mbox{\scriptsize ZTSA}}$. Note that this extra term is a non-polynomial function of $T$, the high temperatures limit of which is     
\begin{equation}\label{s34a} 
\Delta F_{\mbox{\scriptsize free}}(T,L) \xrightarrow[T\gg m]{TL\gg 1}  - \left( \frac{7 A L \pi^2}{180 } \right) T^4    +  \left( \frac{A L m^2}{12 } \right) T^2 - \frac{A L m^4}{8 {\pi}^2}\left[\ln \left( \frac{\pi T}{m}\right) -\gamma \right], 
\end{equation}
where $\gamma \simeq 0.5772$ is the Euler-Maschernoi constant. This expansion can also be obtained by the heat kernel coefficients.

As mentioned above, it has long been recognized that the use of the ZFA yields extra unphysical terms. To remedy this, Geyer et al.~\cite{r17Geyer.} defined a renormalization program in which the polynomial terms obtained using the heat kernel coefficients with powers greater or equal to two are subtracted. In their work on bosonic cases, they emphasized that all of the mentioned terms are of quantum character and do not include the classical term which is proportional to temperature. We now explore the results of this renormalization program. Below, we state the renormalization program as presented in reference~\cite{r17Geyer.},
\begin{equation}\label{s35a}
F^{\mbox{\scriptsize ren}}= E_{0}^{\mbox{\scriptsize ren}} + \Delta_{\mbox{\scriptsize T}} F_{0} 
- \alpha_{0} \left( k_{\mbox{\scriptsize B}} T\right)^4 -\alpha_{1} \left( k_{\mbox{\scriptsize B}} T\right)^3  - \alpha_{2} \left(  k_{\mbox{\scriptsize B}} T\right)^2 . 
\end{equation}
The coefficients of these terms depend on geometrical characteristics of the configuration and can be expressed in terms of heat kernel coefficients.
We calculate these coefficients in Appendix~\ref{appendixF:The Heat kernel method}, and show that they are identical to the those of the high temperature expansions of $F_{\mbox{\scriptsize Zeta}}$, $F_{\mbox{\scriptsize ZTSA}}$, and $\Delta F_{\mbox{\scriptsize free}}$, as given by Eq.~(\ref{s34a}). We also show how the divergent vacuum energy at zero temperature can be obtained by the heat kernel method.
Therefore, based on this renormalization program, the physical Casimir free energy for a massive fermion field confined between two parallel plates obtained using zeta function is as follows 
\begin{equation}\label{s36}
F_{\mbox{\scriptsize Zeta}}^{\mbox{\scriptsize ren}}(T,L)= F_{\mbox{\scriptsize Zeta}} +
      \left( \frac{7 A L \pi^2}{180 } \right) T^4    -  \left( \frac{A L m^2}{12 } \right) T^2 ,
\end{equation}
where $ F_{\mbox{\scriptsize Zeta}}$ is given by Eq.~(\ref{s35d}). One can analogously define a renormalized ZTSA free energy as follows  
\begin{equation}\label{s36b}
F_{\mbox{\scriptsize ZTSA}}^{\mbox{\scriptsize ren}}(T,L) = F_{\mbox{\scriptsize ZTSA}} +
      \left( \frac{7 A L \pi^2}{180 } \right) T^4    -  \left( \frac{A L m^2}{12 } \right) T^2 ,
\end{equation}
where $ F_{\mbox{\scriptsize ZTSA}}$ is given by Eq.~(\ref{s31}).

\begin{center}
\begin{figure}[!h] 
\includegraphics[width=13.5cm]{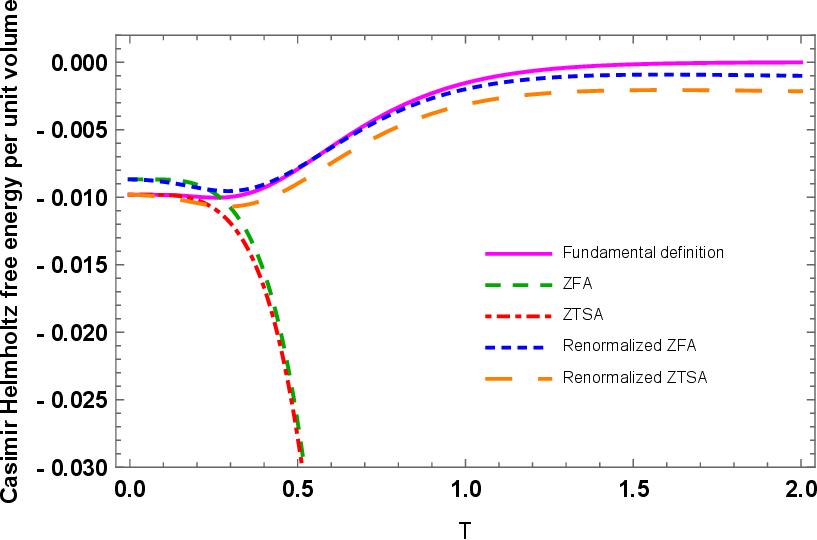}
\caption{\label{fp21} \small
The Casimir free energies per unit volume for a massive fermion field between two parallel plates as a function of temperature with fixed plate separation $L = 1.0$, and mass $m = 0.5$, obtained using five methods within three approaches.}
\end{figure}
\end{center}

To illustrate the differences between the five expressions for the Casimir free energy, we plot them in figure~(\ref{fp21}). As can be seen in this figure, the free energies obtained via the ZFA, and the ZTSA decrease without bound as temperature increases, while the Casimir free energy goes to zero at high temperatures which is an acceptable physical result.
Moreover, the free energies obtained by applying the renormalization program, {\it i.e.}, $F_{\mbox{\scriptsize Zeta}}^{\mbox{\scriptsize ren}}$, $F_{\mbox{\scriptsize ZTSA}}^{\mbox{\scriptsize ren}}$, do not go to zero as temperature increases, and in fact diverge, due to the subtraction of only the first two terms of Eq.~(\ref{s34a}), which are proportional to $T^2$ and $T^4$. The divergence is due to the remaining $\ln(T/m)$ term. The zero temperature limit of both ZTSA results are compatible with that of the  $F_{\mbox{\scriptsize Casimir}}$, while those of the ZFA are not. This is due to the extra temperature-independent term in the $F_{\mbox{\scriptsize Zeta}}$, and $F_{\mbox{\scriptsize Zeta}}^{\mbox{\scriptsize ren}}$, given by the second term in Eq.~(\ref{s35d}). Although it is a common practice to simply neglect any constant term which appear in the results, this temperature-independent term cannot be neglected since it contributes to the Casimir pressure, which is physically measurable.

So far, we have illustrated that the conventional renormalization programs for the ZFA or ZTSA, based on subtracting powers of $T^2$ and higher in the high temperature expansions, do not in general yield the correct results. However, we can still utilize the facility of the zeta function method within the fundamental approach, and devise a new renormalization program. To do this, we need to calculate the free energy of the free massive case by applying the zeta function. We use the first form given by Eq.~(\ref{s4}), integrate over $t$, replace the sum over odd Matsubara frequencies with a sum over all integers minus even integers, which yields
\begin{eqnarray}
F_{\mbox{\scriptsize free}}(T,L)= 4TAL\int\frac{d^3 k}{\left(2 \pi\right)^3}  \lim\limits_{s \to 0} \frac{\partial }{\partial s} \sum\limits_{{n_0} =  1 }^\infty \Bigg\lbrace &\left[ \left(\frac{n_0 \pi}{\beta} \right)^2+M^2\right]^{-s}  - \nonumber\\
&\left[ \left(\frac{2n_0 \pi}{\beta} \right)^2+M^2\right]^{-s}\Bigg\rbrace ,\nonumber 
\end{eqnarray}
where $M^2=k^2+m^2$ and $k^2=K_T^2+k_L^2$. Finally we use the inhomogeneous zeta function to obtain
\begin{eqnarray}\label{s36eew}
&&F_{\mbox{\scriptsize Zeta}}^{\mbox{\scriptsize free}} (T,L) =  \frac{A L}{2\sqrt{\pi^5}} \lim\limits_{s \to 0} \frac{\partial}{\partial s} \int_{0}^{\infty} \frac{dk k^2}{\Gamma (s)} \left[ \omega(k)^{(1-2s)} + \right. \nonumber \\
&&\left. \sum\limits_{j = 1}^{\infty} \left(\frac{\omega(k)}{j \beta}\right)^{\frac{1}{2}-s} \left[2K_{\frac{1}{2}-s} \left(2j \beta \omega(k)\right) -\sqrt{2^{(1-2s )}} K_{\frac{1}{2}-s} \left(j \beta \omega(k)\right) \right]\right].
\end{eqnarray}
Then, using the relationship between the Bessel functions, as given in the footnote of Appendix \ref{appendixB:The Argument Principle}, one can see that $F_{\mbox{\scriptsize Zeta}}^{\mbox{\scriptsize free}}(T,L)$, given by Eq.~(\ref{s36eew}), is equivalent to the sum of the last two terms of $F_{\mbox{\scriptsize Zeta}}(T,L)$, given by Eq.~(\ref{s34b}). This implies that 
\begin{equation}
	F_{\mbox{\scriptsize Casimir}}(T,L)= F_{\mbox{\scriptsize Zeta}}^{\mbox{\scriptsize bounded}}(T,L) - F_{\mbox{\scriptsize Zeta}}^{\mbox{\scriptsize free}}(T,L),
\end{equation}
where we have denoted $F_{\mbox{\scriptsize Zeta}}(T,L)$ by $F_{\mbox{\scriptsize Zeta}}^{\mbox{\scriptsize bounded}}(T,L)$ to emphasis that this is in accord with the fundamental definition. This expression can be looked upon as the correct renormalization scheme, but is nothing more than an, albeit useful, expression for the fundamental definition.

One can now obtain other thermodynamic quantities based on the expressions we have obtained for the free energy using the ZFA given in Eq.~(\ref{s35d}), the ZTSA given in Eq.~(\ref{s31}), and their renormalized versions given in Eqs.~(\ref{s36}, \ref{s36b}). For example, we calculate the pressure for a massive fermion field using the free energy obtained via the zeta function, in analogy with the massless case shown in Eq.~(\ref{s19}). We express the result in terms of $P_{\mbox{\scriptsize Casimir}}(T,L)$, given in Eq.~(\ref{s29}), as follows
\begin{eqnarray}\label{s35}
 P_{\mbox{\scriptsize Zeta}}(T,L) &=&  P_{\mbox{\scriptsize Casimir}}(T,L) -\frac{ m^4}{32 \pi^2}  \left[ 3 - 4 \ln(m) \right]+\Delta P_{\mbox{\scriptsize free}}(T), \hspace{5 mm} \mbox{where} \nonumber \\
\Delta P_{\mbox{\scriptsize free}}(T)&=&-\frac{2 T^2 m^2}{\pi^2} \sum\limits_{j = 1}^{\infty} \frac{(-1)^j}{j^2}  K_{2}\left( j \beta m\right) .
\end{eqnarray}
As before, the second term is a constant term which is a leftover from the use of the zeta function, and the third term is the thermal correction to the pressure of the free case, which the zeta function fails to subtract. 
Next, we calculate the pressure using the free energy obtained via the ZTSA. We express the result in terms of $P_{\mbox{\scriptsize Casimir}}(T,L)$ as follows,
\begin{equation}\label{s61}
P_{\mbox{\scriptsize ZTSA}}(T,L) =  P_{\mbox{\scriptsize Casimir}}(T,L) +\Delta P_{\mbox{\scriptsize free}}(T) .
\end{equation}
Next, we calculate the pressure obtained via the renormalized zeta function, {\it i.e.}, $F_{\mbox{\scriptsize Zeta}}^{\mbox{\scriptsize ren}}(T,L)$,  and the renormalized  ZTSA, {\it i.e.}, $F_{\mbox{\scriptsize ZTSA}}^{\mbox{\scriptsize ren}}(T,L)$. The results are,
\begin{eqnarray}\label{s62}
  P_{\mbox{\scriptsize ZTSA}}^{\mbox{\scriptsize ren}}(T,L) &=&  P_{\mbox{\scriptsize ZTSA}}(T,L)  - \left( \frac{7 \pi^2}{180 } \right) T^4 + \left( \frac{m^2}{12 } \right) T^2 \\
 P_{\mbox{\scriptsize Zeta}}^{\mbox{\scriptsize ren}}(T,L) &=&  P_{\mbox{\scriptsize Zeta}}(T,L)  - \left( \frac{7 \pi^2}{180 } \right) T^4 + \left( \frac{m^2}{12 } \right) T^2   .
 \end{eqnarray}
\begin{center}
	\begin{figure}[h!] 
		\includegraphics[width=13.5cm]{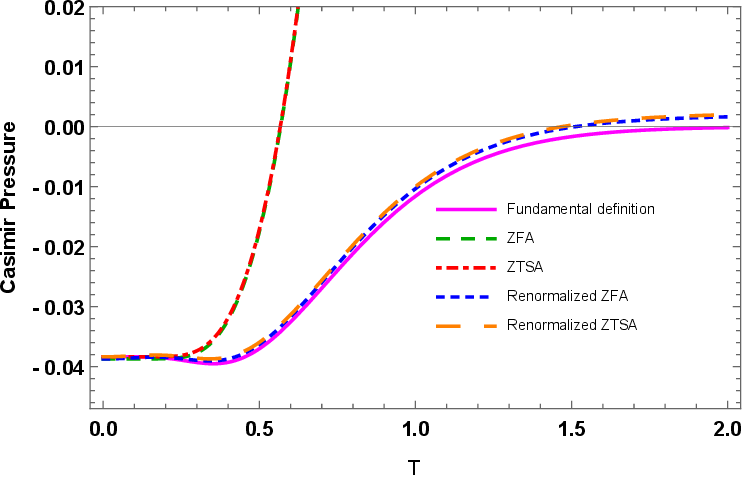}
		\caption{\label{fp50} \small
		The Casimir pressure for a massive fermion field between two parallel plates as a function of temperature with fixed plate separation $L = 1.0$, and mass $m = 0.5$, obtained using five methods within three approaches.}
	\end{figure}
\end{center}
In figure~(\ref{fp50}), we compare these results with the Casimir pressure obtained based on the fundamental approach, given by Eq.~(\ref{s29}). As can be seen, the pressure obtained using the ZFA, and the ZTSA are negative at low temperatures and positive at high temperature, while the Casimir pressure is always negative and goes to zero as temperature increases which is an acceptable physical result. The differences between these results, besides the constant term present in $P_{\mbox{\scriptsize Zeta}}(T,L)$, are due to the thermal correction of pressure of free case $\Delta P_{\mbox{\scriptsize free}}(T)$ which is a non-polynomial function of $T$ for the massive fermion field. The pressure obtained using the ZFA, and the ZTSA all diverge as $T^4$ at high temperatures, and their renormalized versions as $\ln (T/m)$. At $T=0$, only the ZTSA results match the $P_{\mbox{\scriptsize Casimir}}(0,L)$.

On a side note, we can now examine the applicability and limitations of the piston method, which can be used if one is interested only in the Casimir pressure. In this approach, the pressure on the bounded and unbounded sides of the piston are calculated and subtracted. The zeta function method is almost invariably used for this purpose. To trace the cancellations that occur in this subtraction, we first write Eq.~(\ref{s35}), with the labeling mentioned above, as follows 
\begin{eqnarray}\label{pressue}
		P_{\mbox{\scriptsize Zeta}}^{\mbox{\scriptsize bounded}}(T,L) &=&  P_{\mbox{\scriptsize Casimir}}(T,L) + P_{\mbox{\scriptsize Zeta}}^{\mbox{\scriptsize free}}(T), \hspace{1 cm} \mbox{where} \nonumber \\
P_{\mbox{\scriptsize Zeta}}^{\mbox{\scriptsize free}}(T)&=& -\frac{ m^4}{32 \pi^2}  \left[ 3 - 4 \ln(m) \right]+\Delta P_{\mbox{\scriptsize free}}(T),
\end{eqnarray}
Now it is clear that if the zeta function is used to calculate the pressure of both the bounded and unbounded sides of the piston and the results are subtracted, the extra term cancels and one obtains the correct result, {\it i.e.} $P_{\mbox{\scriptsize Casimir}}(T,L)$. In fact, one obtains the correct expression as long as the method used for the two regions is the same, whether it is ZFA, ZTSA or the Abel-Plana formula.


\section{Summary and Discussion}\label{summary}
\indent

In this paper, we have explored the implications and results of the fundamental definition of the Casimir free energy, and how they compare with the results based on two general approaches in common use, {\it i.e.}, the zeta function approach (ZFA), and the zero temperature subtraction approach (ZTSA), including their renormalized versions. We have used the zeta function approach as a representative of the analytic continuation techniques, due to uniqueness of their results. For a concrete example which would illustrate the similarity and differences, we have chosen the massless and massive fermion fields confined between two parallel plates, separated by a distance $L$, with the MIT bag boundary conditions at finite temperature. The fundamental definition of $F_{\mbox{\scriptsize Casimir}}$ is the difference between the free energy of the system in the presence of nonperturbative conditions or constraints, and the one with no constraints, which we have referred to as the free case, both being at the {\em same} temperature $T$ and having the same volume. First, we have calculated the Casimir free energy based on the fundamental definition, and have used it to calculate other Casimir thermodynamic quantities, including the pressure, energy, and entropy, and have shown that all of them are finite and go to zero as the temperature, $L$, or mass increases. This occurs due to the subtraction of the free case at the same temperature, which amounts to the cancellation of both the zero temperature and the thermal correction parts of the bounded case which are equivalent to those of the free case. 
We have also shown that the massless limits of the Casimir thermodynamic quantities obtained for the massive fermion field are identical to the ones obtained for the massless case. 

We have then computed the Casimir thermodynamic quantities using the two other general approaches in common use and compared their results to the ones obtained using the fundamental approach. First, we concentrated on computing the Casimir free energy for a massless fermion field. We first used the zeta function, implemented in three different ways, to evaluate the analytic continuation of the double sums of the spatial and Matsubara modes, and have shown that they all yield equivalent results. Moreover, we have calculated this Casimir free energy using the ZTSA, and have shown that the results are equivalent to those of ZFA. Therefore, we have shown that the use of the ZFA for computing the Casimir free energy of the massless case at finite temperature is equivalent to subtracting the free energy of the free case at zero temperature. However, these results are not equivalent to the ones obtained using the fundamental definition of the Casimir free energy. The difference is the $T^4$ term which is the equivalent of the black-body radiation term. This difference can be traced back to the fact that the free energy of the free case at finite temperature contains the black-body term, the subtraction of which is included in the fundamental definition of the Casimir free energy, but it is not included in the ZFA or the ZTSA. This difference also exists in all other Casimir thermodynamic quantities. For example, $P_{\mbox{\scriptsize Zeta}}$ or $P_{\mbox{\scriptsize ZTSA}}$, contrary to the Casimir pressure obtained by the fundamental approach, are positive at high temperatures, due to the black-body term. A renormalization program has been devised to subtract the high temperature expansions as polynomials in $T$, the use of which yields the correct results for the massless case. 

Next, we have used the ZFA and the ZTSA to calculate the Casimir free energy for a massive fermion field and have shown that, contrary to the massless case, they yield different results. Moreover, similar to the massless case, neither of these results is equivalent to $F_{\mbox{\scriptsize Casimir}}$ obtained via the fundamental definition. The major difference is that both $F_{\mbox{\scriptsize Zeta}}$ and $F_{\mbox{\scriptsize ZTSA}}$ contain the thermal correction to the free case, denoted by $\Delta F_{\mbox{\scriptsize free}}(T,L)$, which they have failed to subtract and is a non-polynomial function of $T$. Moreover, $F_{\mbox{\scriptsize Zeta}}$, includes an extra unphysical temperature-independent term which is a leftover from the use of the inhomogeneous zeta function, is a non-polynomial function of the mass, and goes to zero as $m\to 0$. The high-temperature expansion of $\Delta F_{\mbox{\scriptsize free}}(T,L)$ includes $T^4$, $T^2$ and $\ln (T/m)$ terms. The renormalization program mentioned above removes the first two of these terms in this case, the sum of which does not equal $\Delta F_{\mbox{\scriptsize free}}(T,L)$ for any nonzero temperature. Consequently, $F_{\mbox{\scriptsize Zeta}}^{\mbox{\scriptsize ren}}$ and $F_{\mbox{\scriptsize ZTSA}}^{\mbox{\scriptsize ren}}$ are also not equal to $F_{\mbox{\scriptsize Casimir}}$ at any nonzero temperature. This is in contrast to the massless case, where $\Delta F_{\mbox{\scriptsize free}}(T,L)$ is a simple mononomial $T^4$, the subtraction of which leads to $F_{\mbox{\scriptsize Casimir} }=F_{\mbox{\scriptsize Zeta}}^{\mbox{\scriptsize ren}}=F_{\mbox{\scriptsize ZTSA}}^{\mbox{\scriptsize ren}}$. Therefore, as can be seen from figure~(\ref{fp21}), for the massive case, the five expressions for the Casimir free energy are not equivalent at any temperature, except at $T=0$, where the ZTSA results are equal to $F_{\mbox{\scriptsize Casimir}}(0,L)$. In particular, as $T\to \infty$, $F_{\mbox{\scriptsize Casimir}}\to 0$, $F_{\mbox{\scriptsize Zeta}}$ and $F_{\mbox{\scriptsize ZTSA}}$ $\sim T^4$, and $F_{\mbox{\scriptsize Zeta}}^{\mbox{\scriptsize ren}}$ and $F_{\mbox{\scriptsize ZTSA}}^{\mbox{\scriptsize ren}}$ $\sim \ln(T/m)$. These differences are also present in all other Casimir thermodynamic quantities, e.g., the Casimir pressure illustrated in figure~(\ref{fp50}). 

To summarize, it has long been recognized that the use of the ZFA yields, as does any other approach based on analytic continuation, extra unphysical terms. In particular, in 2008, Geyer et al.~\cite{r17Geyer.} stated that the use of zeta function does not include all necessary subtractions, and they devised a renormalization procedure in which the terms proportional to powers of $T$ higher than the classical terms obtained in the high temperature limit from the heat kernel method, are to be subtracted. Here, we have shown that this approach, as well as the ZTSA, yields the correct results only for the massless cases. We believe that the correct approach for computing the Casimir thermodynamic quantities is by the use of their fundamental definitions. Using this approach, the final results do not contain any extra unphysical terms, and there is no need to devise any new renormalization program, which would even seem impossible for general non-polynomial terms. However, as we have shown, we can utilize the facility of the zeta function in computing the Casimir quantities by their fundamental definitions, {\it e.g.} $F_{\mbox{\scriptsize Casimir} }=F_{\mbox{\scriptsize Zeta}}^{\mbox{\scriptsize bounded}}-F_{\mbox{\scriptsize Zeta}}^{\mbox{\scriptsize free}}$. Here we have denoted $F_{\mbox{\scriptsize Zeta}}$ by $F_{\mbox{\scriptsize Zeta}}^{\mbox{\scriptsize bounded}}$ to emphasize that the zeta function is applied to the bounded case, and this expression for $F_{\mbox{\scriptsize Casimir} }$ is in accordance with its fundamental definition. 
 



\clearpage

\appendix
\numberwithin{equation}{section}

\section{Calculation of the Casimir free energy of the massless case using the Abel-Plana summation formula}
\label{appendixC:FCasimir}
In this appendix, we calculate the Casimir free energy for a massless fermion using its fundamental definition, starting with the second form of the free energy given by Eq.~(\ref{s5}), and show that the final result is equivalent to the result given in Eq.~(\ref{s12}). We first evaluate the integrals over the transverse momenta for both the bounded and free cases using the dimensional regularization, and then subtract the results according to the fundamental definition given by Eq.~(\ref{s6}), to obtain,
\begin{eqnarray}\label{C3}
&&\hspace{-8mm}F_{\mbox{\scriptsize Casimir}} (T,L) = \frac{A \pi^2}{3 L^3}  \left[ \sum\limits_{n_{1} =  0 }^\infty \left(n_1 + \frac{1}{2}\right)^3 - \int_{ 0 }^\infty dk' \left(k'\right)^3\right] +A \sqrt{\frac{8 T^3}{L^3}} \sum\limits_{j=1}^{\infty} \frac{(-1)^j}{\sqrt{j^3}} \times  \nonumber \\
&&\hspace{-8mm} \left[ \sum\limits_{n_{1} =  0 }^\infty \left(n_1 + \frac{1}{2}\right)^{\frac{3}{2}} K_{\frac{3}{2}} \left( \frac{j \pi\left(n_1 + \frac{1}{2}\right)}{TL}\right) - \int_{  0 }^\infty  dk' \left(k'\right)^{\frac{3}{2}} K_{\frac{3}{2}} \left( \frac{j \pi k'}{TL}\right)\right],
\end{eqnarray}
where $k'=k \pi /L$. As can be seen in the above expression, the zero temperature parts of the bounded and free cases, given by the two terms in the first square bracket, are separately divergent, since the expression given in Eq.~(\ref{s5}) contains no analytic continuation. Now, using the Abel-Plana formula (see, for example,~\cite{r43Sahra.})\footnote{The simplest form that is needed here is the following
\begin{equation}
	\sum\limits_{n = 0}^\infty  f\left(n + \frac{1}{2} \right)  = \int_0^\infty f(t)  dt   -
	i \int_0^\infty   \frac{ f(it) - f( - it)}{e^{2 \pi t} + 1} dt. \nonumber
	\end{equation}} the divergences cancel and after simplifying\footnote{Using $\left[(it)^{\frac{3}{2}} K_{\frac{3}{2}} (i t \alpha) - (-it)^{\frac{3}{2}} K_{\frac{3}{2}} (- i t \alpha)\right]= -i \pi t^{\frac{3}{2}} J_{\frac{3}{2}} (t \alpha)$.} we obtain the $F_{\mbox{\scriptsize Casimir}}$ given by Eq.~(\ref{s11}).


We can also obtain another form for the Casimir free energy. First, we expand the logarithm of thermal correction part of the free energy given by Eq.~(\ref{s5}) for large values of $\beta$, then we integrate over the transverse momenta, and finally use the Abel-Plana formula, to obtain,
\begin{equation}\label{C4}
F_{\mbox{\scriptsize Casimir}}(T,L) = 
 -  \frac{7  \pi ^2 A}{2880  L^3}  + \frac{7  \pi ^2  T^4  A L}{180} + \frac{ T^3 A}{\pi } \sum\limits_{j = 1}^\infty  {( - 1)}^j
 \frac{ 1 + \frac{ \pi j}{2 T L} \coth \left( \frac{\pi j}{2 T L} \right)}{j^3  \sinh \left( \frac{\pi j}{2 T L} \right)}  .
\end{equation}
The first term is the zero temperature part and the rest constitute the thermal correction part. This form is equivalent to the result obtained above, {\it i.e.}, Eq.~(\ref{s12}). However, to have an accurate plot using this form, one has keep a large number of terms, otherwise the graph would show an increase from zero at high values of $T$. This is due to the high $\beta$ expansion mentioned above.

\section{Calculation of the free energy using the generalized zeta function}
\label{appendixA:The zeta function}

The most commonly used approach for calculating the Casimir effects is the zeta function approach (ZFA). The generalized zeta function~\cite{r47Elizald.} is given dy the following expression, 
\begin{equation}\label{A1}
Z_p^{M^2}(s ; a_1 ,..., a_p ; c_1 ,..., c_p) = \sum_{n_{1} =  - \infty }^\infty  ...\sum_{n_{p} =  - \infty }^\infty  	{\left[ a_{1} {\left(n_{1} - c_{1}\right)}^2 + ... + a_{p} {\left(n_{p} - c_{p}\right)}^2  + M^2\right]^{ - s}}.
\end{equation}
The above expression yields finite results for $\mathrm{Re} (s) > \frac{p}{2}$, and admits an analytic continuation for $\mathrm{Re} (s) < \frac{p}{2}$,~\cite{r47Elizald., r10Wolf.}. This form is also referred to as the inhomogeneous generalized zeta function. If we set the parameters ${c_1} ,..., {c_p}$ to zero, we obtain a special form of the inhomogeneous generalized zeta function. An important special form called the homogeneous zeta function is obtained when the parameters ${c_1} ,..., {c_p}$, and $M$ are set to zero.  For this case, there is a constraint that the sums should not include the $({n_1}=0, ... ,{n_p}=0)$ mode. Obviously, for the massive case we have to use the inhomogeneous form, while, as shown in the text, both forms can be used for the massless case. 

In the first part of this appendix, we show explicitly three different ways of using the zeta function for obtaining the free energy of the massless case, as outlined in Sec.~\ref{zeta}, starting with Eq.~(\ref{s21}) and obtaining the three equivalent expressions given in Eqs.~(\ref{s23}, \ref{s21bb}, \ref{s21d}). In the first method, we do the double sums simultaneously, so as to obtain the final result shown in Eq.~(\ref{s23}). The expression that we have obtained for $F_{\mbox{\scriptsize Zeta}} (T,L)$, given by Eq.~(\ref{s21}), can be expressed in terms of homogeneous generalized zeta functions as follows, 
\begin{eqnarray}\label{A3a}
&&F_{\mbox{\scriptsize Zeta}} (T,L) = \frac{T A}{4 \pi }  \lim\limits_{s \to 0} \frac{\partial }{\partial s} \frac{\Gamma (s - 1)}{\Gamma (s)} 
\left\lbrace  Z_{2} \left( s - 1 ;  \frac{\pi^2}{\beta^2} , \frac{\pi^2}{4 L^2} \right)   -   \right. \nonumber \\
&& \hspace{-1cm}\left.  Z_{2} \left( s - 1 ; \frac{\pi^2}{\beta^2} , \frac{\pi^2}{ L^2} \right)  -   Z_{2} \left( s - 1 ; \frac{4 \pi^2}{\beta^2} , \frac{\pi^2}{4 L^2} \right)  + Z_{2} \left( s - 1 ; \frac{4 \pi^2}{\beta^2} , \frac{\pi^2}{L^2} \right)    \right\rbrace . \nonumber \\
\end{eqnarray}
Here $ s-1=: s' <1$, and an analytic continuation may be implemented by application of the following zeta function reflection formula~\cite{r12Kris., r48Elizald.,  r482Elizald.},
\begin{equation}\label{A3}
\pi ^{- s'}   \Gamma (s') Z_p \left(s'  ;  {a_1},...,{a_p}\right) = 
\frac{ \pi ^{- \frac{p}{2} + s'} }{\sqrt{{a_1}{a_2}...{a_p}}}  \Gamma (\frac{p}{2} - s') 
Z_p \left( \left(\frac{p}{2} - s'\right) ; \frac{1}{{a_1}},...,\frac{1}{{a_p}}\right).
\end{equation}
Using this for the first term of Eq.~(\ref{A3a}), as an example, we obtain, 
\begin{eqnarray}\label{A3b} 
Z_{2} \left( s - 1 ; \frac{\pi^2}{\beta^2} , \frac{\pi^2}{4 L^2} \right) &=& \frac{2 \beta L \Gamma (2 - s)}{\Gamma (s - 1) {\pi}^{5 - 2s}}  Z_{2} \left( 2 - s; \frac{\beta^2}{\pi^2} , \frac{4 L^2}{\pi^2} \right) \nonumber \\
&=&\frac{2 \beta L \Gamma (2 - s)}{\Gamma (s - 1) {\pi}^{5 - 2s}} \sum\limits_{n_{0} =  - \infty }^\infty  {\sum\limits_{n_{1} =  - \infty }^{\infty '}  \left[ {\left(\frac{{n_0} \beta }{\pi }\right)}^2 + {\left(\frac{{n_1} 2 L}{\pi }\right)}^2\right]^{(s - 2)}}.
\end{eqnarray}
Using the reflection formula for all four terms of Eq.~(\ref{A3a}), taking the derivative with respect to $s$, and taking the limit\footnote{We have used $\lim\limits_{s \to 0} \frac{\partial }{\partial s} \frac{f(s)}{\Gamma (s)} = f(0)$, since $ f(s) $ is an analytic function for $s<1$. } $s \to 0$, the expression for the free energy becomes
\begin{eqnarray}\label{A3c} 
F_{\mbox{\scriptsize Zeta}} (T,L) &=& - \frac{7 A L}{ 2 \pi^2 } \sum\limits_{{n_0} =  1 }^\infty \frac{1}{({n_0} \beta)^4}  -  \frac{7 A L}{ 32 \pi^2 } \sum\limits_{{n_1} =  1}^\infty \frac{1}{({n_1} L)^4}  + \nonumber\\
&&\frac{A L}{ 2 \pi^2 } \sum\limits_{{n_0} = 1 }^\infty  \sum\limits_{{n_1} = 1}^{\infty}  \Bigg[  4  {\left( n_0^2 \beta^2 + 4 n_1^2 L^2 \right)}^{- 2} - 2 {\left( n_0^2 \beta^2 +  n_1^2 L^2\right)}^{- 2} \nonumber\\
&&\left.     - 2 {\left( \frac{n_0^2 \beta^2}{4} + 4 n_1^2 L^2\right)}^{- 2}  + {\left( \frac{n_0^2 \beta^2}{4} +  n_1^2 L^2\right)}^{- 2}     \right] .
\end{eqnarray}
Since the summations in Eq.~(\ref{A3c}) are over only positive definite integers, we use the homogeneous form of the generalized inhomogeneous Epstein zeta function~\cite{r48Kris2.}, given by 
\begin{equation}\label{A4}
{E_p^{M^2}}(s ; {a_1} ,..., {a_p}) = \sum_{n_{1} =  1 }^\infty  {...\sum_{n_{p} = 1}^\infty  {\left[ a_1 {n_1}^2 + a_2 {n_2}^2 + ... + a_p {n_p}^2  + M^2\right]}^{ - s}}     .
\end{equation}
That is, we use $E_p^{0}$ which is usually denoted by $E_p$. Before we apply this to the four terms in Eq.~(\ref{A3c}), we use the following relation for the Epstein zeta function, $E_2$,  
 \begin{eqnarray}\label{A6}
E_2(s;{a_1},{a_2}) =   - \frac{\zeta (2s)}{2 {a_1}^{s}} + \sqrt {\frac{\pi }{{a_2}}}  
\frac{\Gamma (s - \frac{1}{2}) \zeta (2s - 1)}{2  \Gamma (s)  {a_1}^{\left(s - \frac{1}{2} \right)}} + \nonumber \\ 
\frac{2 \pi^s}{\sqrt{{a_2}^{\left( s + \frac{1}{2} \right)}} \Gamma (s) \sqrt{{a_1}^{\left( s - \frac{1}{2} \right)}}}   \sum\limits_{{m_1} = 1}^\infty  {\sum\limits_{{m_2} = 1}^\infty  {\left( \frac{{m_2}}{{m_1}} \right)}^{\left(s - \frac{1}{2}\right)} K_{ \frac{1}{2} - s} \left( 2 \pi {m_1}{m_2} \sqrt{\frac{{a_1}}{{a_2}}} \right)}.
\end{eqnarray}
Using this for each term in Eq.~(\ref{A3c}), and computing the sum over $n_0$ modes\footnote{We have used the following identities,\\ $\sum\limits_{m = 1}^\infty  \sqrt {m^3}  K_{\frac{3}{2}}(ma) = 	\sqrt {\frac{\pi }{ 2 a^3}} \frac{\left(a + 1\right)  e^a - 1}{{\left( e^a - 1 \right)}^2} = \sum\limits_{m = 1}^\infty  \sqrt {m^3}  K_{ - \frac{3}{2}}(ma)$.}, we obtain
\begin{eqnarray}\label{A6b} 
&&F_{\mbox{\scriptsize Zeta}} (T,L) = - \frac{7 A L T^4 \pi^2 }{ 180 } + \frac{A T}{8 \pi L^2} \sum\limits_{{n_1} = 1}^{\infty} \frac{1}{n_1^3}  \left[   5  \frac{ \left( 4 \pi {n_1} L T + 1\right) e^{4 \pi {n_1} L T } - 1 }{{\left(e^{4 \pi {n_1} L T } - 1\right)}^2} - \right. \nonumber \\
&& \left.    4  \frac{ \left( 2 \pi {n_1} L T + 1\right) e^{2 \pi {n_1} L T } - 1 }{{\left(e^{2 \pi {n_1} L T } - 1\right)}^2}   - \frac{ \left( 8\pi {n_1} L T + 1\right) e^{8 \pi {n_1} L T } - 1 }{{\left(e^{8 \pi {n_1} L T } - 1\right)}^2}     \right]  .
\end{eqnarray}
Then, we simplify the above expression and obtain the free energy given by Eq.~(\ref{s23}).

Next, we compute the free energy of the massless case using the zeta function to do the sums separately. To do this, first we note that there are partial cancellations in the sums of Eq.~(\ref{s21}), {\it i.e.}, the terms with $n_0 =0$ or $n_1 =0$ cancel each other. Next, we express the remaining sums as sums over positive integers.

For our first case, which constitutes our second method, we first calculate the sum over spatial modes, and then the sum over the remaining Matsubara modes. To do this, we consider the Matsubara modes, {\it i.e.}, $n_0 \pi /\beta$ and $2n_0 \pi /\beta$, as the constant term of Eq.~(\ref{A4}). Then, we use the following expression for $E_1^{M^2}(s ; a)$ ~\cite{r48Kris2.} 
\begin{eqnarray}\label{A8}
E_1^{M^2}(s ; a) &=& -  \frac{1}{2 M^{2s}}  +
\sqrt {\frac{\pi }{a}}   \frac{1}{2 \Gamma{(s)}  M^{2s - 1}}
\Bigg[ \Gamma{(s - \frac{1}{2})} +  \nonumber \\
&&\left. 4 \sum_{j = 1}^\infty  {\left( \frac{\sqrt{ a}}{\pi j M} \right)}^{\left( \frac{1}{2} - s\right)}  K_{\frac{1}{2} - s}{\left(\frac{2 \pi j M}{\sqrt{a}}\right)}\right],
\end{eqnarray}
to compute the free energy. Then, we obtain
\begin{eqnarray}\label{A9}
&&\hspace{-8mm}F_{\mbox{\scriptsize Zeta}} (T,L) = \frac{T A L}{2\sqrt{\pi^3} }  \lim\limits_{s \to 0} \frac{\partial }{\partial s} \frac{1}{\Gamma (s)} 
\sum\limits_{{n_0} = 1 }^\infty {\left( \frac{{n_0} \pi}{\beta} \right)}^{3 - 2 s}  
\Bigg\{  \Gamma\left(s - \frac{3}{2} \right) \left(1- 2^{3-2s}\right) +   \nonumber \\
&&\hspace{-8mm} 4 \sum\limits_{{n_1} = 1 }^\infty {\left( \frac{\beta }{n_0 {n_1} L \pi} \right)}^{\frac{3}{2} - s}\left[ \left( 2^{s - \frac{1}{2}} + 2^{ \frac{3}{2} - s}\right) K_{ \frac{ 3}{2} - s} \left( 4 \pi {n_0}{n_1} L T \right) - \right. \nonumber \\
&&\hspace{-2mm}\left.  K_{ \frac{ 3}{2} - s} \left( 2 \pi {n_0}{n_1} L T \right) -2  K_{ \frac{ 3}{2} - s} \left( 8 \pi {n_0}{n_1} L T \right) \right]     \Bigg\} .
\end{eqnarray}
Taking the derivative with respect to $s$ and the limit $s \to 0$, except for the first term which includes a divergent sum, the free energy becomes
\begin{eqnarray}\label{A10}
&&\hspace{-8mm}F_{\mbox{\scriptsize Zeta}} (T,L) =  - \frac{7 T A L}{2\sqrt{\pi^3} }  \lim\limits_{s \to 0} \frac{\partial }{\partial s} \frac{\Gamma\left(s - \frac{3}{2} \right)}{\Gamma (s)}  \sum\limits_{{n_0} =  1 }^\infty  \left( \frac{{n_0} \pi}{\beta} \right)^{3 - 2 s} + A \sqrt{\frac{T^5}{ L }} \sum\limits_{{n_0} = 1 }^\infty \times \nonumber \\
&&\hspace{-8mm}  \sum\limits_{{n_1} = 1}^{\infty} \sqrt{\frac{n_0^3}{n_1^3}}  \Bigg\{ 5 \sqrt{2}  K_{ \frac{- 3}{2}} \left( 4 \pi {n_0}{n_1} L T \right) - 2  K_{ \frac{- 3}{2}} \left( 2 \pi {n_0}{n_1} L T \right)    - 4  K_{ \frac{- 3}{2}} \left( 8 \pi {n_0}{n_1} L T \right)     \Bigg\} . \nonumber \\
\end{eqnarray}
This expression is equivalent to Eq.~(\ref{s21b}) and, as mentioned in Sec.~\ref{zeta}, we calculate the divergent sum over the Matsubara frequencies using the analytic continuation obtained via $\zeta(-3)$. 

For our second case, which constitutes our third method, we first calculate the sum over Matsubara modes and then the sum over the remaining spatial modes. To do this, we consider the spatial modes, {\it i.e.}, $n_1 \pi / 2L$ and $n_1 \pi /L$, as the constant term of Eq.~(\ref{A4}). Then, we use Eq.~(\ref{A8}) and compute the free energy, obtaining the following expression 
\begin{eqnarray}\label{A11}
&&\hspace{-8mm}F_{\mbox{\scriptsize Zeta}} (T,L)=  -\frac{7 A}{32\sqrt{\pi^3} }  \lim\limits_{s \to 0} \frac{\partial }{\partial s} \frac{\Gamma\left(s - \frac{3}{2} \right)}{\Gamma (s)}  \sum\limits_{{n_1} =  1 }^\infty \left( \frac{{n_1} \pi}{L} \right)^{3 - 2 s}  +A \sqrt{\frac{T^3}{ L }} \sum\limits_{{n_0} = 1 }^\infty  \times \nonumber \\
 &&\hspace{-8mm} \sum\limits_{{n_1} = 1}^{\infty} \sqrt{\frac{n_1^3}{n_0^3}}  \left\lbrace   \frac{5}{\sqrt{2}}  K_{ \frac{- 3}{2}} \left(\frac{ \pi {n_0}{n_1}}{ L T} \right)  
 - 2  K_{ \frac{- 3}{2}} \left( \frac{ 2 \pi {n_0}{n_1}}{ L T}\right)    -  K_{ \frac{- 3}{2}} \left(\frac{ \pi {n_0}{n_1}}{2 L T} \right)     \right\rbrace  .
\end{eqnarray}
This expression is equivalent to Eq.~(\ref{s21c}) and, as mentioned in Sec.~\ref{zeta}, we calculate the divergent sum over the spatial modes using the analytic continuation rendered by $\zeta(-3)$, which yields the correct zero temperature part present in the expression for our final result given by Eq.~(\ref{s21d}).

In the last part of this appendix, we use the inhomogeneous generalized zeta functions to compute the free energy of the massive case.
As mentioned in Sec.~\ref{Epstein}, we start with the first form of the free energy given by Eq.~(\ref{SZeta}), and we present the sum over Matsubara frequencies as the difference between sum over all integers and even integers, as used in Eq.~(\ref{s21}), to obtain
\begin{eqnarray}\label{A12}
F_{\mbox{\scriptsize Zeta}} (T,L) &=& \frac{T A}{\pi }  \lim\limits_{s \to 0} \frac{\partial }{\partial s} \frac{\Gamma (s - 1)}{\Gamma (s)} 
\sum\limits_{{n_0} =  1 }^\infty \sum\limits_{{n_1}>0 }  \left\lbrace  \left[ \left(\frac{{n_0} \pi }{\beta }\right)^2 + k_{n_1}^2 + m^2\right]^{1 - s}  - \right. \nonumber \\
&&\left.\left[ \left(\frac{2 {n_0} \pi }{\beta }\right)^2 + k_{n_1}^2 + m^2\right]^{1 - s}   \right\rbrace  .
\end{eqnarray}
To calculate the sum over Matsubara modes, we consider the mass term and the irregular spatial modes in Eq.~(\ref{A12}), {\it i.e.}, $k_{n_1}^2 + m^2$, as the constant term of Eq.~(\ref{A4}), {\it i.e.}, $M^2$. Then, we use Eq.~(\ref{A8}) to obtain the free energy given by Eq.~(\ref{s33}).  

\section{Calculation of the summation over irregular modes using the Principle of the Argument theorem}
\label{appendixB:The Argument Principle}

The Principle of the Argument theorem relates the difference between the number of zeros and poles of a meromorphic function $f(z)$, to a contour integral of the logarithmic derivative of the function~\cite{r49Ahlf.}. In this paper, we use the generalized form of the Principle of the Argument theorem which is as follows~\cite{r49Ahlf.}
\begin{equation}\label{B1}
\sum_{n} g(a_n) - \sum_{m} g(b_m) = \frac{1}{2 \pi i} \oint_C g(z)  d\left[ \ln(f(z)) \right],
\end{equation}
where ${a_n}$ and ${b_m}$ are the zeroes and poles of $f(z)$ inside the closed contour $C$, respectively, and $g(z)$ is assumed to be an analytic function in the region enclosed by the contour $C$. In applying this theorem to our problem, we find it convenient to use the following generalization of Eq.\ (\ref{B1})
\begin{equation}\label{B2}
\sum_{n} g(a_n) - \sum_{m} g(b_m) = \frac{1}{2 \pi i} \oint_C g(z)  d\left[ \ln(f(z)  h(z))\right],
\end{equation}
with the condition that the function $h(z)$ should be analytic and have no zeros in the region enclosed by the contour $C$.

\begin{figure}[h!] 
\centering
\includegraphics[scale=.5]{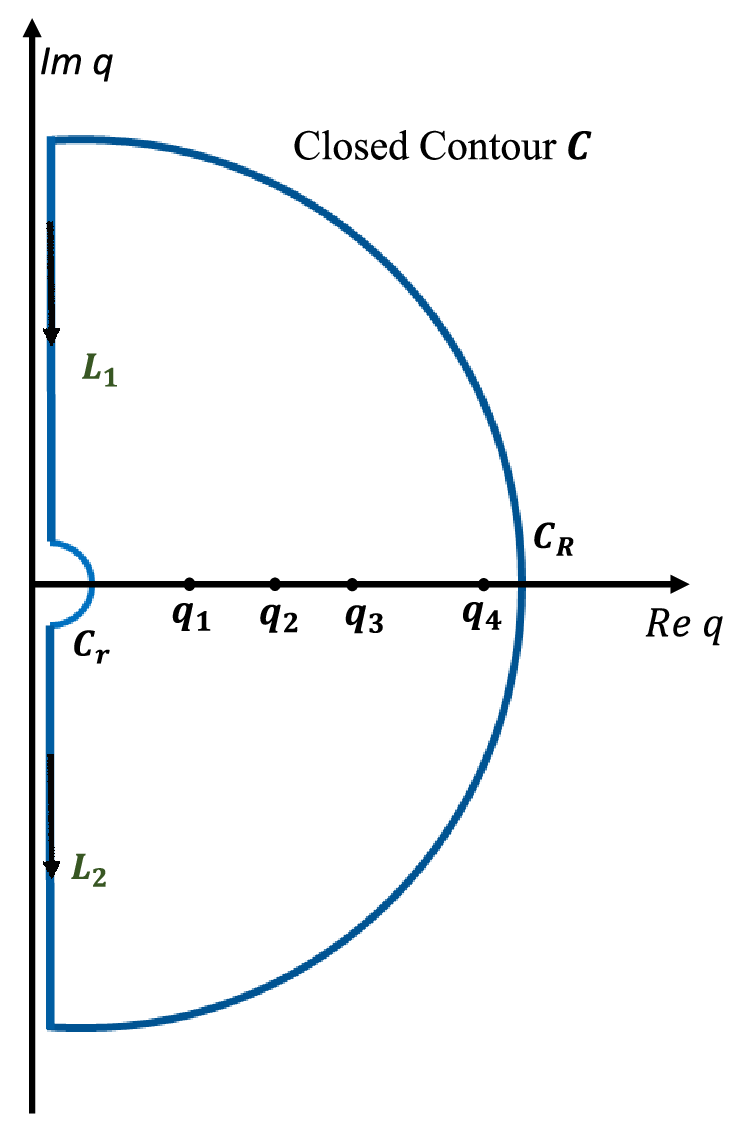}
\caption{\label{fp46a} \small
The closed integration contour $C$, referred to in Eq.~(\ref{B3}), for evaluating the free energy of the massive fermion between two parallel plates, where $q_i$ is related to the irregular spatial mode by $q_i^2 = k_i^2 + m^2$. As $R \to \infty $ and $r \to 0$, only $L_1$ and $L_2$ give nonzero contributions.}
\end{figure}

The expression that we have obtained for the free energy of the massive case between two plates in Sec.~\ref{massive}, using its fundamental definition and given by Eq.~(\ref{s24}), contains a sum over the irregular spatial modes which are the roots of $f(k_{n_1})$ in Eq.~(\ref{s1000}). We use the Principle of the Argument theorem, as expressed in Eq.~(\ref{B2}), to compute this sum and obtain 
\begin{eqnarray}\label{B3}
&&\hspace{-8mm}F_{\mbox{\scriptsize bounded}}(T,L)   = \frac{A}{\sqrt{\pi^3}} \frac{1}{2 \pi i}  \oint_{C}   \lim\limits_{s \to 0} \frac{\partial}{\partial s} \frac{1}{\Gamma (s)} \left[\frac{\Gamma\left( s - \frac{3}{2}\right)}{4} q^{3 - 2s}
+  \right.\nonumber \\
&&\left. \sum\limits_{{n_0} = 1}^\infty  (-1)^{n_0}   \left(\frac{2 q}{ n_0 \beta}\right)^{\frac{3}{2} - s} 
 K_{\frac{3}{2} - s}\left(\beta {n_0} q\right)  \right] \times \nonumber\\
&&\hspace{-8mm}  d\left\{\ln\left[ \frac{2\sqrt{q^2 - m^2}  \cos\left(\sqrt{q^2 - m^2} L \right) +2 m  \sin\left(\sqrt{q^2 - m^2} L \right)}{\sqrt{q^2 - m^2} + i  m}\right]\right\},  
\end{eqnarray}
where $g(q_{n_1}^2) = g({k_{n_1}^2} + m^2)$ is the summand in Eq.~(\ref{s24}), while $g(q)$ is the integrand defined in Eq.~(\ref{B2}). We have chosen $h(q) = 2  / ( \sqrt{q^2 - m^2} + i m)$. 
The closed contour $C$ in the complex $q$-plane should enclose all of the roots of $f(k_{n_1})$. As can be seen in figure~(\ref{fp46a}), the closed contour $C$ is composed of two arcs, $C_{R}$ and  $C_{r}$, and also two straight line segments $L_{1}$, and $L_{2}$. To compute this contour integral over $q$, we replace the first term in the integrand in Eq.~(\ref{B3}), {\it i.e.}, the $q^{3 - 2s}$ term, by the following integral representation
\begin{eqnarray}\label{B3ab}
q^{3 - 2s} = \int_{0}^{\infty} \frac{e^{- t q^2}dt}{\sqrt{t^{5-2s}} \Gamma \left(s - \frac{ 3}{2}\right)}.
\end{eqnarray}
Next, we integrate by parts. In the limit $R \to \infty $ and $r \to 0$, only $L_1$ and $L_2$ give nonzero contributions, which can be written as follows 
\begin{eqnarray}\label{B4}
\hspace{-8mm} F_{\mbox{\scriptsize bounded}}(T,L)   =  \frac{i A}{\sqrt{\pi^5} }\lim\limits_{s \to 0} \frac{\partial}{\partial s}   \int_{- i \infty}^{i \infty}  dq  \left[ \int_{0}^{\infty} \frac{dt  e^{- t q^2} q}{4 \sqrt{t^{3-2s}} \Gamma (s) }   
+  \sum\limits_{{n_0} = 1}^\infty (-1)^{n_0}  \left(\frac{2 q}{n_0 \beta}\right)^{\frac{1}{2} - s}  \times   \right.  \nonumber\\
\hspace{-2mm}  \left. \frac{q K_{\frac{1}{2} - s}\left(\beta {n_0} q\right)}{\Gamma (s)} \right] \left\{\ln\left[ \frac{2\sqrt{q^2 - m^2}  \cos\left(\sqrt{q^2 - m^2} L \right) +2 m  \sin\left(\sqrt{q^2 - m^2} L \right)}{\sqrt{q^2 - m^2} + i  m}\right]\right\}.  \nonumber \\
\end{eqnarray}
After changing variable $q = ip$, and evaluating the integral over $t$, we express the results as follows
\begin{eqnarray}\label{B5}
&&\hspace{-6mm}F_{\mbox{\scriptsize bounded}}(T,L)   = -  \frac{A}{\sqrt{ \pi^5} } \lim\limits_{s \to 0} \frac{\partial}{\partial s} \frac{1}{\Gamma (s)}  \int_{0}^{ \infty}  dp  \Bigg\{  \frac{\Gamma \left( s - \frac{1}{2} \right)}{4}\left[ \left(i p\right)^{2 - 2s} +  \left(- i p\right)^{2 - 2s} \right]+    \nonumber\\
&&\hspace{-6mm} \sum\limits_{{n_0} = 1}^\infty (-1)^{n_0}  \left(\frac{2}{n_0 \beta}\right)^{\frac{1}{2} - s} \left[ \left(i p\right)^{ \frac{3}{2} - s} K_{ \frac{1}{2} - s} \left( i p \beta n_0\right) + \left(- i p\right)^{ \frac{3}{2} - s} K_{ \frac{1}{2} - s} \left(- i p \beta n_0\right) \right] \Bigg\}  \times \nonumber\\
&&\hspace{-6mm} \left\{ \ln\left[ e^{L \sqrt{p^2 + m^2} } \left(1+ \frac{ \sqrt{p^2 + m^2} - m}{ \sqrt{p^2 + m^2} + m} e^{- 2 L \sqrt{p^2 + m^2} }\right)\right] \right\}.  
\end{eqnarray} 
We can simplify\footnote{Using $\sqrt{(i p)^{3 - 2s}} K_{\frac{1}{2} - s} \left(i p a \right) +\sqrt{(- i p)^{3 - 2s}} K_{\frac{1}{2} - s} \left(- i p a \right) = \pi \sqrt{p^{3 - 2s}} J_{\frac{1}{2} - s} \left(p a \right) $.} this expression to obtain the free energy for the bounded case given by Eq.~(\ref{s1003}).

In Sec.~\ref{Epstein}, we have calculated the free energy of a massive fermion using the inhomogeneous zeta function and have displayed the result in Eq.~(\ref{s34}). The details of calculations are as follows. We start with Eq.~(\ref{s33}), follow the same steps as above, and use the same contour shown in figure~(\ref{fp46a}). We obtain
\begin{eqnarray}\label{B7}
&&\hspace{-8mm} F_{\mbox{\scriptsize Zeta}}(T,L) = \frac{i A}{ \sqrt{\pi^5} }   \int_{- i \infty}^{i \infty}  dq  \left[ \int_{0}^{\infty} \frac{dt  e^{- t q^2} q}{ 4 t^{\frac{3}{2} - s} }   
+   \right.  \nonumber\\
&& \hspace{-8mm}  \left.  2 \sum\limits_{{n_0} = 1}^\infty   {\left(\frac{q}{n_0 \beta}\right)}^{\frac{3}{2} - s}  \left( K_{\frac{1}{2} - s}\left(2\beta {n_0} q\right) - \sqrt{2^{\left(1+ 2 s\right)}} K_{\frac{1}{2} - s}\left(\beta {n_0} q\right) \right)\right] \times  \nonumber\\
&& \hspace{-8mm}  \left\{\ln\left[ \frac{2\sqrt{q^2 - m^2}  \cos\left(\sqrt{q^2 - m^2} L \right) +2 m  \sin\left(\sqrt{q^2 - m^2} L \right)}{\sqrt{q^2 - m^2} + i  m}\right]\right\}.
\end{eqnarray}
Using the change of variable $q=i p$, these integrals are converted to the Euclidean form. Next, we evaluate the integral over $t$ and simplify the resulting expression to obtain the expression for $F_{\mbox{\scriptsize Zeta}}$ given by Eq.~(\ref{s34}).

\section{Calculation of the Casimir free energy using the Dimensional Regularization}
\label{appendixD:The Dimensional Regularization}

In this appendix, we calculate the Casimir free energy for a massive fermion field, based on its fundamental definition, using the second form of the free energy given by Eq.~(\ref{s5}). For the bounded region, we evaluate the integral over the transverse momenta using the dimensional regularization, and obtain 
\begin{eqnarray}\label{D1}
F_{\mbox{\scriptsize bounded}}(T,L) &=& 2 A \sum\limits_{n_1 >0} \lim\limits_{D \to 2} \left[ \frac{ \Gamma \left(-\frac{D+1}{2}\right)}{\left(4 \pi\right)^{\frac{D+1}{2}}} \omega_{n_1}^{D+1} + \right.\nonumber \\
&&\left.  4\sum\limits_{j=1}^{\infty} (-1)^j \left(\frac{\omega_{n_1}}{2 \pi \beta j}\right)^{\frac{D+1}{2}} K_{\frac{D+1}{2}} \left( \beta j \omega_{n_1}\right) \right], 
\end{eqnarray}
where $\omega_{n_1}=\sqrt{k_{n_1}^2+m^2}$.
Then, we evaluate the sum over the irregular spatial modes using the Principle of the Argument theorem, which is the same procedure as done for the bounded case in Sec. \ref{massive} given by Eq.~(\ref{s24}) (see Appendix \ref{appendixB:The Argument Principle}), and obtain
\begin{eqnarray}\label{D3}
&&\hspace{-8mm}F_{\mbox{\scriptsize bounded}}(T,L) = -8A \lim\limits_{D \to 2} \int_0^{\infty} dp \left[ \frac{p^D}{2 \Gamma \left(\frac{D+1}{2}\right) \left(4 \pi\right)^{\frac{D+1}{2}}} +\sum\limits_{j=1}^{\infty} (-1)^j \times \right.  \nonumber \\
&&\hspace{-8mm}\left. \left(\frac{p}{2 \pi \beta j}\right)^{\frac{D+1}{2}} J_{\frac{D-1}{2}} \left( \beta j p\right) \right] \left[ L \omega(p) + \ln \left(1 + \frac{\omega(p) - m}{\omega(p) + m} e^{- 2 L \omega (p)}\right) \right], 
\end{eqnarray}
where $\omega(p)=\sqrt{p^2+m^2}$. Then, we calculate the integral over $p$ for the first part of the above expression, which does not include the logarithm function, and simplify the resulting expression. Then, the free energy for the bounded case becomes
\begin{eqnarray}\label{D3a}
&&F_{\mbox{\scriptsize bounded}}(T,L)= \lim\limits_{D \to 2}\left[ \frac{2AL m^{D+2} \Gamma \left( -\frac{D+2}{2}\right)}{\left(4 \pi\right)^{\frac{D+2}{2}}} + 8AL \sum\limits_{j=1}^{\infty} (-1)^j \left(\frac{m}{2 \pi \beta j}\right)^{\frac{D+2}{2}} \times \right.\nonumber \\
&&\left. K_{\frac{D+2}{2}} \left( \beta j m\right) - 4A\int_{0}^{\infty} dp \ln \left(1 + \frac{\omega(p) - m}{\omega(p) + m} e^{- 2 L \omega (p)}\right) \times\right.  \nonumber\\
&& \left.  \left[ \frac{p^D}{2 \Gamma \left(\frac{D+1}{2}\right) \left(4 \pi\right)^{\frac{D+1}{2}}} +\sum\limits_{j=1}^{\infty} (-1)^j \left(\frac{p}{2 \pi \beta j}\right)^{\frac{D+1}{2}} J_{\frac{D-1}{2}} \left( \beta j p\right) \right]  \right].
\end{eqnarray}
Next, we calculate the free energy of the free case at finite temperature, by starting with the second form of the free energy, and using dimensional regularization to calculate the integrals over momenta and obtain 
\begin{eqnarray}\label{D4}
F_{\mbox{\scriptsize free}}(T,L) &=& \lim\limits_{D \to 2}\left[ \frac{2AL m^{D+2} \Gamma \left( -\frac{D+2}{2}\right)}{\left(4 \pi\right)^{\frac{D+2}{2}}} +  \right.\nonumber \\
&&\left. 8AL \sum\limits_{j=1}^{\infty} (-1)^j \left(\frac{m}{2 \pi \beta j}\right)^{\frac{D+2}{2}} K_{\frac{D+2}{2}} \left( \beta j m\right)\right] . 
\end{eqnarray}
As can be seen, the first two terms of $F_{\mbox{\scriptsize bounded}}(T,L)$, given by Eq.~(\ref{D3a}), are identical to the two terms of $F_{\mbox{\scriptsize free}}(T,L)$, given by Eq.~(\ref{D4}). 
Notice that the first terms actually diverge as $\Gamma \left( -2\right)$, in contrast to the analogous terms in Eqs.~(\ref{s1003bv}, \ref{s25}) which are obtained using the first form of the free energy Eq.~(\ref{s4}), which has an embedded analytic continuation.
After subtracting these terms, and taking the limit $D \to 2$, we obtain the same expression for the Casimir free energy as in Eq.~(\ref{s27}).

\section{Calculation of the Casimir free energy for massive fermions using the Boyer method}
\label{appendixE:The Boyer method}

\begin{figure}[h!] 
	\centering
	\includegraphics[scale=.4]{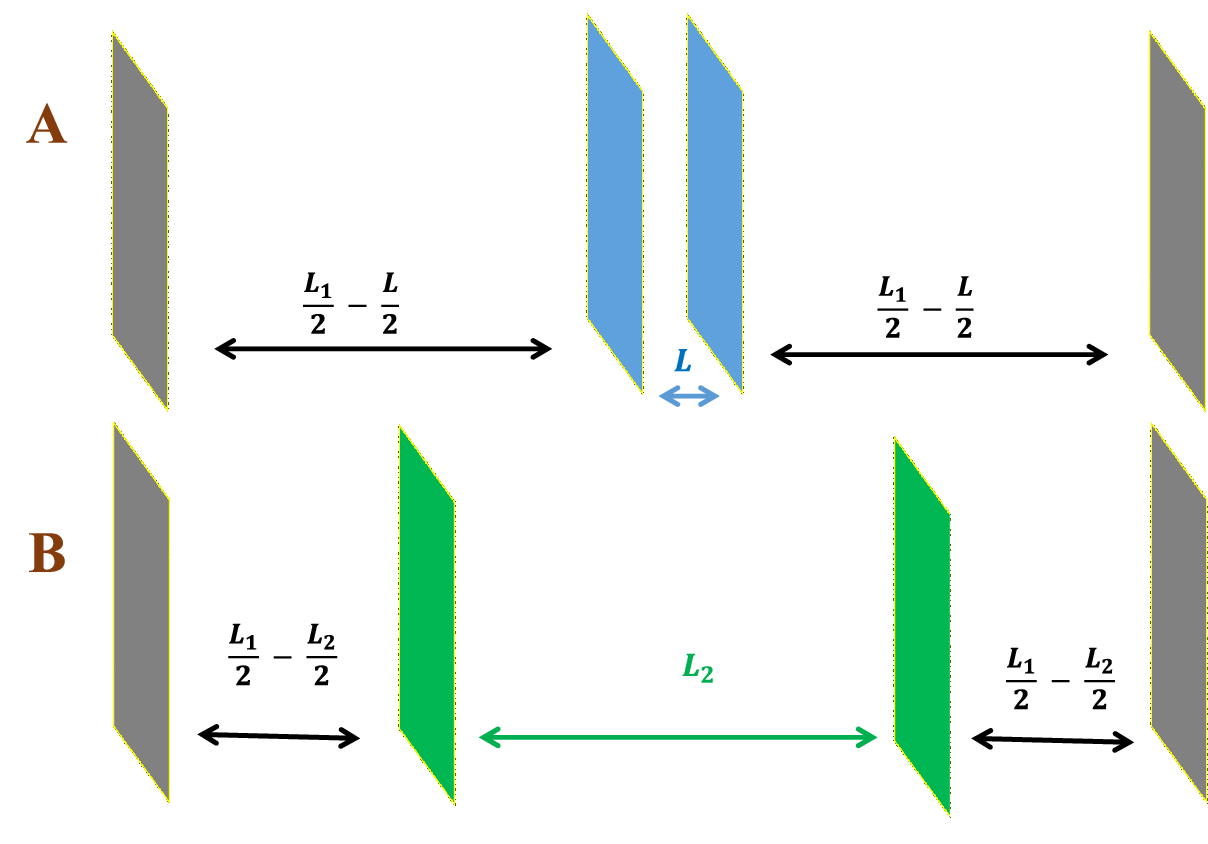}
	\caption{\label{fp100} \small
		The geometry of the two different configurations whose free energies for a massive fermion field are to be subtracted. In the upper (lower) configuration, labeled $A$ ($B$), two inner plates are located at $z=\pm L/2$ ($z=\pm L_2/2$) surrounded by two outer plates located  at $z=\pm L_1 /2$. We impose the MIT boundary conditions on all plates.}
\end{figure}

In this appendix, we calculate the Casimir free energy for a massive fermion field using the Boyer method~\cite{r50Boyer.} and show that the final result is equivalent to the result given in Eq.~(\ref{s27}), obtained using the fundamental approach. In this method, we subtract the free energies of two configurations, at the same temperature, and obtain the Casimir free energy of our original system by taking appropriate limits. Configuration $A$ consists of two inner plates located at $z=\pm L/2$ surrounded by two outer plates located at $z=\pm L_1 /2$. Configuration $B$ is similar to $A$ except the two inner plates are located $z=\pm L_2 /2$, with $ L< L_2 <L_1$, as depicted in  figure~(\ref{fp100}).  

The Casimir energy can be defined in terms of the difference between the free energies of configurations $A$ and $B$ as follows
\begin{eqnarray}\label{E1}
F_{\mbox{\scriptsize Casimir}}(T,L) =\lim\limits_{L_2 \to \infty} \Big\lbrace \lim\limits_{L_1 \to \infty} \left[ F_A(T,L,L_1) - F_B(T,L_2,L_1)\right] \Big\rbrace ,
\end{eqnarray}
where $F_A(T,L,L_1) = F_{\mbox{\scriptsize bounded}}^{I}(T,L)+2 F_{\mbox{\scriptsize bounded}}^{II}(T,L,L_1)$ and $F_B(T,L_2,L_1) = F_{\mbox{\scriptsize bounded}}^{I}(T,L_2)+2 F_{\mbox{\scriptsize bounded}}^{II}(T,L_2,L_1)$. Moreover, $F_{\mbox{\scriptsize bounded}}^{I}$ denotes the free energy between two inner plates and the $F_{\mbox{\scriptsize bounded}}^{II}$ denotes the free energy of bounded regions adjacent to the inner plates.
In fact, the Boyer method can be thought of as a rigorous implementation of the fundamental definition, provided the two configurations are taken to be at the same temperature, in which any possible contributions from the regions outside of the bounded region is also taken into account.


To calculate the free energy for each of six regions shown in figure~(\ref{fp100}), we use the result obtained in the Sec.~\ref{massive} for the $F_{\mbox{\scriptsize bounded}}$, given by Eq.~(\ref{s1003bv}). For example, the free energy for the outer bounded regions of Configuration $B$ becomes
\begin{eqnarray}\label{E2}
&&F_{\mbox{\scriptsize bounded}}^{II}(T,L_2,L_1) =  \frac{A (L_1 - L_2)}{4\sqrt{\pi^5}} \lim\limits_{s \to 0} \frac{\partial}{\partial s}\frac{ \Gamma \left(s - \frac{1}{2}\right)}{\Gamma (s)} \int_{ 0}^{ \infty}  p^{2 - 2s}  \omega(p) dp	+ \nonumber \\
&&\frac{A (L_1 - L_2) T^2 m^2}{\pi^2} \sum\limits_{n_0 = 1}^{\infty} \frac{(-1)^{n_0}}{n_0^2} K_2 \left(n_0 \beta m\right) - \frac{A}{\pi^2} \int_{0}^{\infty} \left[ p^2 + \right. \nonumber \\
&& \left. 2T \sum\limits_{n_0 = 1}^{\infty} \frac{(-1)^{n_0}}{n_0} p \sin \left(n_0 \beta p\right)\right]  \ln\left( 1 + \frac{\omega(p)  -   m}{\omega(p)  + m} e^{ - (L_1 - L_2) \omega(p)}\right)  dp  .
\end{eqnarray} 
As mentioned before, only the first term of the above expression contains a divergent part. Moreover, $F_{\mbox{\scriptsize bounded}}^{II}(T,L_2,L_1)=F_{\mbox{\scriptsize bounded}}^{II}(T,L_1-L_2)$ and the first two terms are linear in $L_1-L_2$. Adding the contributions of the three regions of configuration $B$ we obtain
\begin{eqnarray}\label{E3}
&&F_B(T,L_2,L_1) =  \frac{A L_1}{2\sqrt{\pi^5}} \lim\limits_{s \to 0} \frac{\partial}{\partial s}\frac{ \Gamma \left(s - \frac{1}{2}\right)}{\Gamma (s)} \int_{ 0}^{ \infty}  p^{2 - 2s}  \omega(p) dp	+\frac{2 A L_1 T^2 m^2}{\pi^2} \times  \nonumber \\
&&\sum\limits_{n_0 = 1}^{\infty} \frac{(-1)^{n_0}}{n_0^2} K_2 \left(n_0 \beta m\right) - \frac{A}{\pi^2} \int_{0}^{\infty} dp \left[ p^2 + 2T \sum\limits_{n_0 = 1}^{\infty} \frac{(-1)^{n_0}}{n_0} p \sin \left(n_0 \beta p\right)\right]  \nonumber \\
&& \left[ \ln\left( 1 + \frac{\omega(p)  -   m}{\omega(p)  + m} e^{ - 2 L_2 \omega(p)}\right) +2\ln\left( 1 + \frac{\omega(p)  -   m}{\omega(p)  + m} e^{ - (L_1 - L_2) \omega(p)}\right) \right]   .
\end{eqnarray} 
We can compute $F_A(T,L,L_1)$ similarly. Upon using Eq.~(\ref{E1}) to calculate $F_{\mbox{\scriptsize Casimir}}(T,L)$, the first two terms of $F_A(T,L,L_1)$ and $F_B(T,L_2,L_1)$, which include divergent integrals, cancel even before we take the limits and we obtain
\begin{eqnarray}\label{E4}
&&F_{\mbox{\scriptsize Casimir}}(T,L) = - \frac{A}{\pi^2} \int_{0}^{\infty} dp \Bigg\lbrace p^2 + 2T \sum\limits_{n_0 = 1}^{\infty} \frac{(-1)^{n_0}}{n_0} p \sin \left(n_0 \beta p\right)\Bigg\rbrace \times  \nonumber \\
&& \hspace{-9mm} \lim\limits_{L_2 \to \infty}  \left[ \lim\limits_{L_1 \to \infty} \left[ \ln\left( 1 + \frac{\omega(p)  -   m}{\omega(p)  + m} e^{ - 2 L \omega(p)}\right) - \ln\left( 1 + \frac{\omega(p)  -   m}{\omega(p)  + m} e^{ - 2 L_2 \omega(p)}\right) + \right. \right.   \nonumber \\
&&  \hspace{-9mm} \left.  \left. 2 \ln\left( 1 + \frac{\omega(p)  -   m}{\omega(p)  + m} e^{ - (L_1 - L) \omega(p)}\right) -2\ln\left( 1 + \frac{\omega(p)  -   m}{\omega(p)  + m} e^{ - (L_1 - L_2) \omega(p)}\right) \right] \right].
\end{eqnarray}
Finally, upon taking the limits $L_1 \to \infty$, and $L_2 \to \infty$, sequentially, we obtain the same expression for the Casimir free energy given by Eq.~(\ref{s27}). This proves that, when using the fundamental approach, the contributions from the outer regions in the bounded case are precisely canceled by the corresponding contributions of the free case.

\section{The heat kernel coefficients}
\label{appendixF:The Heat kernel method}

The heat kernel expansion is an important tool in the computations of the Casimir effects, which can be used to obtain the high and low temperature limits of the Casimir thermodynamics quantities, including the divergences in the vacuum energy~\cite{r31Bord2., r12Kris., r17Geyer.}. In the first part of this appendix, we obtain the divergent terms of the energy at zero temperature for our model by calculating the nonzero heat kernel coefficients. To obtain the energy, we use the partition function at zero temperature for a massive free fermion in path integral representation:
\begin{equation}\label{F1}
Z[0] = \int  D \overline \psi  D \psi \exp\left\{- i \int {d^4 x} \left[ \overline \psi \left( i \slashed{\partial} -m\right) \psi \right] \right\}  =  \Det \left(i \slashed{\partial}  - m \right).
\end{equation}
Using the effective action, the vacuum energy for time-independent boundaries and backgrounds is obtained as \cite{r31Bord2.}
\begin{equation}\label{F2}
E =  \frac{i}{T } \Ln(Z[0]) =   \frac{i}{T } \Ln\left[\Det \left(i \slashed{\partial} - m \right)\right]= 
\frac{2 i}{T}  \Tr \left[ \Ln\left( -P^{2} + m^2 \right) \right],
\end{equation}
where $T$ is the total time and the trace indicates the summation over eigenvalues of Dirac operator in the momentum space representation. The explicit form of the energy at zero temperature is
\begin{eqnarray}\label{F3}
&&\hspace{-9mm}E _{\mbox{\scriptsize bounded}}(0,L)= -\frac{2i}{ T} \int_{-\infty}^{\infty} \frac{Td\omega}{2 \pi} \int \frac{Ad^2 K_T}{\left(2 \pi \right)^2}  \sum\limits_{n_1} \lim\limits_{s \to 0}   \frac{\partial }{\partial s}  \left[ - \omega^2 + \omega _{{n_1} , {K_T}}^2 \right]^{-s}\nonumber \\
&&\hspace*{+9mm} = -2i \int_{-\infty}^{\infty} \frac{d\omega}{2 \pi} \int \frac{Ad^2 K_T}{\left(2 \pi \right)^2}  \sum\limits_{n_1} \lim\limits_{s \to 0}   \frac{\partial }{\partial s} \int_{0}^\infty \frac{dt e^{-t\left( - \omega^2 + \omega _{{n_1} , {K_T}}^2\right)}}{t^{1-s}\Gamma (s)},
\end{eqnarray}
where $\omega_{{n_1} , {K_T}}=\sqrt{K_T^2+ k_{n_1}^2+m^2}$. Due to the MIT boundary conditions at the plates for the massive case, the longitudinal momentum $k_{n_1}$ takes on discrete values which are solutions to Eq.~(\ref{s1000}).
Note that the second form of energy given by Eq.~(\ref{F3}) has an embedded analytic continuation, as mentioned in Sec.~\ref{Helmholtz free energy}. After performing a wick rotation on $\omega$, we evaluate its integral and, to obtain the nonzero heat kernel coefficients for our model at zero temperature, present the result in terms of the heat kernel of the spatial part as 
\begin{equation}\label{F4}
\hspace{-9mm} E _{\mbox{\scriptsize bounded}}(0,L)= \lim\limits_{s \to 0} \frac{\partial }{\partial s} \frac{1}{\Gamma (s)}\int_{0}^\infty \frac{dt e^{ -tm^2}}{\sqrt{\pi}t^{\frac{3}{2} - s}} \mathbf{K}(t),
\end{equation}
where the heat kernel for our model is the following
\begin{equation}\label{F5}
\mathbf{K}(t)=\int \frac{Ad^2 K_T}{\left(2 \pi \right)^2} \sum\limits_{n_1} e^{ - t \left(K_T^2+k_{n_1}^2\right) } .
\end{equation}
The general form for the expansion of the heat kernel is \cite{r31Bord2.},
\begin{equation}\label{F6}
\hspace{-5mm}	\mathbf{K}(t) = \sum\limits_J e^{-t \Lambda_J} =\frac{1}{(4\pi t)^{\frac{3}{2}}} \sum\limits_{n=0}^\infty a_{\frac{n}{2}} t^{\frac{n}{2}}.
\end{equation}
The integrand in Eq.~(\ref{F4}) has a pole at $t=0$. So, to obtain the divergent part of this integral, we divide the interval of the integration into $t \in [0, 1]$ and $t \in [1,\infty)$. Then we need to only evaluate the integral over $t$ in the first interval, for which we use the expansion of $\mathbf{K}(t)$ and the exponential mass term as follows:
\begin{equation}\label{F7}
\mathbf{K}(t) e^{-t m^2} =\frac{1}{(4\pi t)^{\frac{3}{2}}}\sum\limits_{j=0}^\infty \alpha_{\frac{j}{2}} t^{\frac{j}{2}}=\sum\limits_{j=0}^\infty \left(\sum\limits_{k=0}^{[\frac{j}{2}]} (-1)^k  \frac{m^{2k}}{k!}  a_{\frac{j}{2}-k}\right)  t^{\frac{j}{2}}, 
\end{equation}
where $ a_{j/2-k}$ are the heat kernel coefficients of the spatial part. Then we obtain 
 \begin{equation}\label{F8}
\hspace{-9mm} E _{\mbox{\scriptsize bounded}}(0,L)= \frac{1}{8 \pi^2} \lim\limits_{s \to 0} \frac{\partial }{\partial s} \frac{1}{\Gamma (s)} \sum\limits_{j=0}^\infty \left[\frac{\alpha_{\frac{j}{2}}}{s-2+\frac{j}{2}}\right].
\end{equation}
The expression in the bracket in Eq.~(\ref{F8}) has a simple pole for $j=4$. Therefore, the energy in terms of the only nonzero expansion coefficient, {\it i.e.}, $\alpha_2$, becomes
\begin{equation}\label{F9}
\hspace{-9mm} E _{\mbox{\scriptsize bounded}}(0,L)= \frac{1}{16 \pi^2} \lim\limits_{s \to 0} \frac{\partial }{\partial s} \frac{1}{\Gamma (s)}  \left[\frac{2 a_2+2 a_1 m^2+ a_0 m^4}{s}\right].
\end{equation}
Now, we obtain the heat kernel coefficients of the spatial part, {\it i.e.}, $a_{\frac{n}{2}}$, which include a sum of two local integrals, one over the volume and the other over the surface~\cite{r31Bord2.}. According to our model which includes two separate plates with MIT bag model boundary condition, the surface part of plates cancels each other and only the volume part gives a nonzero contribution. So, the only nonzero coefficient for this model is $a_0=AL$. Hence, the divergence of the integral part of expression for the energy at zero temperature, given by Eq.~(\ref{F4}), can be inferred from Eq.~(\ref{F9}) and is $\frac{ALm^4}{16 \pi^2s}$. 
To compare this expression with the analogous one that we have obtained for the bounded case of massive fermion, shown in the footnote of Sec.~\ref{massive}, {\it i.e.}, $ [- m^{4-2s}/(4\sqrt{\pi})] \Gamma(3/2-s)  \Gamma(s-2)$, and also the divergent integral of $F_{\mbox{\scriptsize Zeta}}$ in Eq.~(\ref{s34b}), we expand this terms for $s \to 0$ and obtain
\begin{eqnarray}\label{F10}
&&\hspace{-9mm} F _{\mbox{\scriptsize bounded}}^{\mbox{\scriptsize div. part}}(0,L) =- \frac{AL}{2\sqrt{\pi^5}} \lim\limits_{s \to 0} \frac{\partial }{\partial s} \frac{1}{\Gamma (s)}  \left[\frac{m^{4-2s} \Gamma \left(s-\frac{1}{2}\right) \Gamma \left(\frac{3}{2}-s\right)\Gamma \left(s-2\right)}{4 \sqrt{\pi}}\right]\nonumber \\
&&\hspace*{15mm}=- \frac{AL}{8 \pi^3} \lim\limits_{s \to 0} \frac{\partial }{\partial s} \frac{1}{\Gamma (s)}\left[-\frac{\pi m^4}{2s} +\frac{m^4}{4}\left( 4\pi \ln(mL) -3\pi\right)\right].
\end{eqnarray}
As can be seen, the first term of the bracket in Eq.~(\ref{F10}) is equivalent to the corresponding divergent contribution in terms of the heat kernel coefficients.

In the last part of this appendix, we obtain the nonzero heat kernel coefficients for our model at high temperatures. We first, present Eq.~(\ref{s4}) in terms of the heat kernel of the spatial part, and express the sum over Matsubara frequencies as the difference between sum over all integers and even integers to obtain
\begin{equation}\label{F11}
F_{\mbox{\scriptsize bounded}}(T,L) = -\frac{4}{\beta} \lim\limits_{s \to 0} \frac{\partial}{\partial s} \sum\limits_{n_0 = 1}^\infty \int_0^\infty \frac{dt e^{-t m^2} }{\Gamma (s) t^{1-s}} \mathbf{K}(t) \left[e^{-t \left(\frac{n_0 \pi}{\beta}\right)^2} - e^{-t \left(\frac{2 n_0 \pi}{\beta}\right)^2}\right],
\end{equation}
where the $\mathbf{K}(t)$ is given by Eq.~(\ref{F5}) for our model. Next, we evaluate the integral over $t$ using the expansion of $e^{-t m^2}\mathbf{K}(t)$ given by Eq.~(\ref{F7}), and obtain 
\begin{eqnarray}\label{F12}
\hspace{-9mm} F(T,L) &=& \frac{T}{2\sqrt{\pi^3}} \sum\limits_{j=0}^{\infty}  \lim\limits_{s \to 0} \frac{\partial}{\partial s} \left(\frac{\pi}{\beta}\right)^{3-2s-j} \mathbf{\alpha}_{\frac{j}{2}} \Gamma \left(\frac{2s-3+j}{2}\right) \times\nonumber\\
&& \frac{\zeta (2s+j-3)}{\Gamma (s) } \left(1 -2^{3-2s-j}\right).
\end{eqnarray}
After evaluating $\lim\limits_{s \to 0} \frac{\partial}{\partial s}$ and simplifying, we obtain the following expression for the high temperature expansion of the free energy
\begin{eqnarray}\label{F13}
\hspace{-9mm}F(T,L) &\underset{\mathrm{\beta \ll 1}}{\longrightarrow}& - \frac{7 \pi^2 \mathbf{\alpha}_0}{180}  T^4   - \frac{3 \zeta (3) \mathbf{\alpha}_{\frac{1}{2}}}{4 \sqrt{\pi^3}} T^3 - \frac{\mathbf{\alpha}_1}{12} T^2 -\frac{\ln (2) \mathbf{\alpha}_{\frac{3}{2}}}{2 \sqrt{\pi^3}} T +\nonumber\\
&& \frac{\mathbf{\alpha}_2}{4 \pi^2}\left(\gamma - \ln (\pi T)\right) +... .
\end{eqnarray} 
As mentioned above, the only nonzero heat kernel coefficient of the spatial part for our model is $a_0$. So, one can easily obtain the heat kernel coefficients that appear in Eq.~(\ref{F13}) at the high temperature limit and express them as follows
\begin{eqnarray}\label{F14}
\mathbf{\alpha}_0=AL,   \quad\quad  \mathbf{\alpha}_{\frac{1}{2}}=\mathbf{\alpha}_{\frac{3}{2}}=0,    \quad\quad   \mathbf{\alpha}_1=- m^2 AL,  \quad\quad  \mathbf{\alpha}_2=\frac{m^4}{2} AL . 
\end{eqnarray}
Comparing these results with the asymptotic expansion of the thermal correction to the free energy, given in Eq.~(\ref{s34a}), one observes that the extra unphysical terms which appear in both $F_{\mbox{\scriptsize Zeta}}$ and $F_{\mbox{\scriptsize ZTSA}}$ at high temperatures are equivalent to the terms obtained using the heat kernel coefficients at high temperatures.

\end{document}